\documentclass{aa}   

\bibpunct{(}{)}{;}{a}{}{,} 
\usepackage{graphicx}
\usepackage{txfonts}
\usepackage{empheq}
\graphicspath{{images//}}
\usepackage{subcaption}
\usepackage[pdftex, hidelinks]{hyperref}

\let\orgautoref\autoref
\providecommand{\Autoref}
{\def\equationautorefname{Equation}%
  \def\figureautorefname{Figure}%
  \def\subfigureautorefname{Figure}%
  \def\Itemautorefname{Item}%
  \def\tableautorefname{Table}%
  \def\sectionautorefname{Section}%
  \def\subsectionautorefname{Section}%
  \def\subsubsectionautorefname{Section}%
  \def\chapterautorefname{Section}%
  \def\partautorefname{Part}%
\orgautoref}

\renewcommand{\autoref}
{\def\equationautorefname{Eq.}%
  \def\figureautorefname{Fig.}%
  \def\subfigureautorefname{Fig.}%
  \def\Itemautorefname{item}%
  \def\tableautorefname{Table}%
  \def\sectionautorefname{Sect.}%
  \def\subsectionautorefname{Sect.}%
  \def\subsubsectionautorefname{Sect.}%
  \def\chapterautorefname{Sect.}%
  \def\partautorefname{Sect.}%
\orgautoref}

\begin{document} 

\title{A high-performance and portable asymptotic preserving radiation hydrodynamics code with the M\textsubscript{1} model}

\author{H. Bloch     \inst{1}\thanks{\email{helene.bloch@cea.fr}}
  \and P. Tremblin   \inst{1}
  \and M. Gonz\'alez \inst{2,3}
  \and T. Padioleau  \inst{1}
  \and E. Audit      \inst{1}
}

\institute{Universit\'e Paris-Saclay, UVSQ, CNRS, CEA, Maison de la Simulation, 91191, Gif-sur-Yvette, France\
  \and Universit\'e de Paris, AIM, F-91191 Gif-sur-Yvette, France\
  \and AIM, CEA, CNRS, Universit\'e Paris-Saclay, F-91191 Gif-sur-Yvette, France
}

\date{Received 04 June 2020; accepted 22 November 2020}

\abstract{}
{We present a new radiation hydrodynamics code, called "ARK-RT" which uses a two-moment model with the M\textsubscript{1} closure relation for radiative transfer. This code aims at being ready for high-performance computing, on exascale architectures.}
{The two-moment model is solved using a finite volume scheme. The scheme is asymptotic preserving to capture accurately both optically thick and thin regimes. We also propose a well-balanced discretization of the radiative flux source term able to capture constant flux steady states with discontinuities in opacity. We use the library Trilinos for linear algebra and the package Kokkos allows us to reach high-performance computing and portability across different architectures, such as multi-core, many-core, and GP-GPU.}
{ARK-RT is able to reproduce standard tests in both free-streaming and diffusive limits, including purely radiative tests and radiation hydrodynamics ones. Using a time-implicit solver is profitable as soon as the time step given by the hydrodynamics is $50-100$ times larger than the explicit time step for radiative transfer, depending on the preconditioner and the architecture. Albeit  more work is needed to ensure stability in all circumstances. Using ARK-RT, we study the propagation of an ionization front in convective dense cores. We show that the ionization front is strongly stable against perturbations even with destabilizing convective motions. As a result, the presence of instabilities should be interpreted with caution. Overall, ARK-RT is well-suited to study many astrophysical problems involving convection and radiative transfer such as the dynamics of \ion{H}{ii} regions in massive pre-stellar dense cores and future applications could include planetary atmospheres.}
{}

\keywords{radiative transfer --
  hydrodynamics --
  methods: numerical
}

\maketitle

\section{Introduction} \label{sect:intro}

In many astrophysical situations, radiation is an important process which interacts with the surrounding gas, e.g. in (exo) planet's atmospheres (e.g. \citealt{thomas2002}), massive stars (e.g. \citealt{kuiper2010, migonerisse2020}), \ion{H}{ii} regions (e.g. \citealt{spitzer1978}), up to the cosmic reionization (e.g. \citealt{massimo2009}). In all these situations, the radiation can be absorbed, thus heating the surrounding gas. Photons could also be emitted by the gas or by another source. Photons could be scattered, which will change their direction of propagation and perhaps their frequency \citep{chandrasekhar1960}. 

Two main regimes can arise, depending on the mean free path of photons compared to the characteristic length of the system \citep{mihalas1984}. On one hand, in the diffusive limit, the medium is optically thick (mean free path of photons much smaller than the characteristic length), the radiation and the matter strongly interact with each other. On the other hand, in the free-streaming regime, the radiation does not affect the gas, and the medium is optically thin (mean free path of photons greater than the characteristic length). Numerically, it is difficult to accurately capture both limits. 

Due to the high number of degrees of freedom (i.e. the time, the position, the direction of propagation, and the frequency of photons), only a few problems can be solved analytically \citep{chandrasekhar1960} and direct simulations are out of reach for modern computers. Different models have been developed to reduce the computational cost. We will focus on the moment models \citep{levermore1984}: the specific intensity is averaged over the direction of propagation of photons. It presents several advantages: the computational cost is lower than other methods such as Monte-Carlo method and, mostly, it is easy to couple it with a grid-based hydrodynamics code.

One can consider only the moment of order $0$ (the radiative energy), leading to the flux-limited diffusion (FLD) approximation \citep{levermore1981}. Because this model considers only the moment of order $0$, its computational cost is quite low, but it is very diffusive in the free-streaming regime. To tackle this issue, one can use a two-moment model (radiative energy and radiative flux), with the M\textsubscript{1} closure relation \citep{dubroca1999}. However, this method can suffer from artifacts when multiple beams cross in the free-streaming regime \citep{gonzalez2006}. One can solve this issue by using a three-moment model (radiative energy, radiative flux, and radiative pressure) with the M\textsubscript{2} closure relation \citep{pichard2016}. However, because of the increase of unknowns, the computational cost also increases. In this work, we have chosen to use the two-moment model with the M\textsubscript{1} closure relation because the computational cost remains affordable, and we do not encounter in our applications the problem of beams crossing in the free-streaming regime.

Even though the M\textsubscript{1} model is accurate in both free-streaming and diffusive regimes at the continuous level, numerical schemes also need to properly capture both limits. Several approaches have been developed. For example, \citet{berthon2011} presented a scheme based on an HLL solver with source terms modified with a free parameter. Following this idea, we propose a new so-called asymptotic preserving scheme, also based on an HLL solver. Nevertheless, we have chosen another parameter to recover the asymptotic behavior, in the diffusive limit. Furthermore, our integration of source terms is different. In many physical applications (e.g. clouds), optically thick regions are found next to optically thin zones. We propose a well-balanced modification of the source term, which allows us to accurately reach steady states in the presence of sharp transitions.

As a first application, we are interested in the development of \ion{H}{ii} regions in massive pre-stellar dense cores \citep{churchwell2002}. We focus on solving the radiation hydrodynamics equations. An explicit solver for the radiative transfer would be restricted by a Courant-Friedrichs-Lewy (CFL) condition, limited by the speed of light. This will result in a very low time step compared to the hydrodynamics one, which is limited by the speed of sound of the fluid. Several methods have been developed to get around this problem. The one we have chosen is a time-implicit solver (e.g. \citealt{gonzalez2007}). The temporality of the radiative transfer will be preserved, which is not the case with the reduced speed of light approximation (RSLA, e.g. \citealt{gnedin2001}). However, this method is costly because it requires solving large sparse linear systems.

Fortunately, linear algebra for high-performance computing (HPC) has been investigated over the years. Because the linear system we will have to solve is large and sparse, direct methods are out of reach. Iterative solvers with preconditioners have been developed to tackle this issue (e.g. \citealt{saad2003}). In this work, we use the library Trilinos \citep{heroux2005} because it allows us to target different architectures, such as multi-core, many-core, and GP-GPU. It also provides, among others, algebraic multigrid (AMG) preconditioners.

The paper is organized as follows. In the next section, we present more precisely the moment model and the M\textsubscript{1} closure relation. We go through our new numerical scheme, well-balanced and asymptotic preserving in the diffusive limit in \autoref{sect:numericalScheme}. In \autoref{sect:implementationAndParallelization}, we give details of some implementation features of our code ARK-RT, especially the Kokkos and Trilinos libraries used for shared memory parallelism and linear algebra for high-performance computing. We also show some performance results. In \autoref{sect:numericalResults}, we present some numerical test cases to show the importance of the asymptotic preserving and well-balanced properties. Finally, in \autoref{sect:hiiRegionExpansion} we present a physical application about the stability of the ionization front in \ion{H}{ii} region dense cores.

\section{Model} \label{sect:model}

\subsection{Physical model} \label{sect:physicalModel}

In this work, we only consider gray radiative transfer, i.e. we have computed the average over the frequency. We also assume local thermodynamic equilibrium (LTE) and we do not consider scattering. The mathematical description of radiative transfer was formalized by \citet{chandrasekhar1960}:
\begin{equation}
  \left( \frac{1}{c} \frac{\partial}{\partial t} + \vec{\Omega} \cdot \nabla \right) I(\vec{x}, t, \vec{\Omega}) = \sigma B(\vec{x}, t) - \sigma I(\vec{x}, t, \vec{\Omega}),
  \label{eq:specificIntensity}
\end{equation}
where $I$, the quantity of interest, is the specific intensity, $c$ is the speed of light, $\sigma$ is the opacity of the medium and $B$ is the black body specific intensity. $\vec{x}$, $t$, and $\vec{\Omega}$ are the spatial, temporal, and angular variables. This model can be generalized to multigroup radiative transfer (e.g. \citealt{turpault2005}).

Because the specific intensity $I$ depends on six variables in the three-dimensional case, the numerical treatment of \autoref{eq:specificIntensity} becomes rapidly costly. To reduce the computational cost, we use a moment method, detailed in the next section.

\subsection{M\textsubscript{1} model} \label{sect:m1Model}

Let us consider the three first moments of the specific intensity: the radiative energy $E_r$, the radiative flux $\vec{F}_r$, and the radiative pressure $\tens{P}_r$ defined as:
\begin{equation}
  \begin{alignedat}{4}
    E_r &=& \frac{1}{c} &\oint_{4\pi} I(\vec{x}, t, \vec{\Omega}) d\vec{\Omega} \\
    \vec{F}_r &= && \oint_{4\pi} I(\vec{x}, t, \vec{\Omega}) \vec{\Omega} d\vec{\Omega} \\
    \tens{P}_r &=& \frac{1}{c} & \oint_{4\pi} I(\vec{x}, t, \vec{\Omega}) \vec{\Omega} \otimes \vec{\Omega} d\vec{\Omega}.
  \end{alignedat}
\end{equation}
The mean over solid angles of \autoref{eq:specificIntensity} and its product by $\vec{\Omega}$ give the following system:
\begin{subequations}
  \begin{empheq}[left=\empheqlbrace]{align}
    \partial_t E_r + \nabla \cdot \vec{F}_r  &  = \begin{aligned}[t]
      & c \sigma \left( a_r T_g^4 - E_r \right) \\\label{eq:momentModel:energy}\\
    \end{aligned}\\
    \partial_t \vec{F}_r + c^2 \nabla \cdot \tens{P}_r &= -c \sigma \vec{F}_r \label{eq:momentModel:flux},
  \end{empheq}
  \label{eq:momentModel}
\end{subequations}
where $T_g$ is the gas temperature and $a_r$ is the radiation constant.

The fluid and the radiation exchange energy and momentum through emission and absorption. To ensure the conservation of the total energy when the hydrodynamics is frozen, the energy exchange term is given by
\begin{equation}
  \partial_t (\rho c_v T_g) = - c \sigma \left( a_r T_g^4 - E_r \right).
  \label{eq:internalEnergyEqua}
\end{equation}
$\rho c_v T_g$ is the gas internal energy, with $\rho$ the density of the fluid and $c_v$ the heat capacity, defined by $c_v = \frac{k_b}{\mu m_H (\gamma-1)}$ for a perfect gas, where $k_b$ is the Boltzmann constant, $\mu$ is the mean molecular weight, $m_H$ is the mass of hydrogen and $\gamma$ is the adiabatic index of the gas.

We still have to specify a closure relation, i.e. a way to express $\tens{P}_r$ as a function of $E_r$ and $\vec{F}_r$. The one chosen here is the M\textsubscript{1} model \citep{levermore1984}. From \citet{dubroca1999}, we write $\tens{P}_r = \tens{D} E_r$ where $\tens{D}$ is the Eddington tensor, defined by $\tens{D} = \frac{1-\chi}{2} \tens{I} + \frac{3\chi-1}{2} \vec{n} \otimes \vec{n}$ with $\chi$ the Eddington factor, $\tens{I}$ the identity matrix and $\vec{f} = \frac{\vec{F}_r}{c E_r} = f \vec{n}$ is the reduced flux. $\chi$ can be specified either by applying a Lorentz transform to an isotropic distribution of photons \citep{levermore1984}, either by minimizing the radiative entropy \citep{dubroca1999}. In both cases, we have $\chi = \frac{3+4f^2}{5+2\sqrt{4-3f^2}}$. Let us notice that $f \le 1$, which ensures that the radiative energy cannot be transported faster than the speed of light.

The M\textsubscript{1} model preserves both free-streaming and diffusive limits. On one hand, if $f = 1$, then $\tens{P}_r = E_r \vec{n} \otimes \vec{n}$, only the transport regime remains. On the other hand, if $f = 0$, the model in the diffusive regime simplifies into the P\textsubscript{1} model, with $\tens{P}_r = \frac{1}{3} E_r \tens{I}$. The radiative pressure tensor becomes isotropic. We will look into the diffusion limit more precisely in \autoref{sect:diffusiveLimit}.

\subsection{Radiation hydrodynamics} \label{sect:rhd}

We now consider the radiation hydrodynamics equations. The fluid evolution is described by the Euler equations expressing conservation of mass, balance of momentum, and balance of energy. Because photons are relativistic particles, we have to evaluate the quantities in the laboratory frame or the comoving frame, which is moving with the fluid. On one hand, using the comoving frame introduces non-conservative terms in the left-hand side of the equations. On the other hand, the hyperbolic part of the system remains simple in the laboratory frame, but some source terms have to be incorporated to describe the interactions between the matter and the radiation. We have chosen the second approach. These source terms characterize the momentum and energy exchanges between the fluid and the radiation: 
\begin{subequations}
  \begin{empheq}[left=\empheqlbrace]{align}
    \partial_t \rho + \nabla \cdot (\rho \vec{u})   &= \begin{aligned}[t]
      0\label{eq:radiativeHydrodynamicsEqua:mass}\\
    \end{aligned}\\
    \partial_t (\rho \vec{u}) + \nabla \cdot (\rho \vec{u} \otimes \vec{u} + p \tens{I}) &=  \rho \vec{g} + \frac{\sigma}{c} \vec{F}_r - \frac{1}{c} \vec{S}_{\vec{F}_r}(\vec{u})\label{eq:radiativeHydrodynamicsEqua:momentum}\\
    \notag \partial_t (\rho E) + \nabla \cdot ((\rho E+p) \vec{u}) &=  \rho \vec{g} \cdot \vec{u} \\
    &-c \sigma \left( a_r T_g^4 - E_r \right) - S_{E_r}(\vec{u}) \label{eq:radiativeHydrodynamicsEqua:energy}\\
    \partial_t E_r + \nabla \cdot \vec{F}_r &=   c \sigma (a_r T_g^4 - E_r) + S_{E_r}(\vec{u}) \label{eq:radiativeHydrodynamicsEqua:radiativeEnergy}\\
    \partial_t \vec{F}_r + c^2 \nabla \cdot \tens{P}_r &=   - c \sigma \vec{F}_r + c \vec{S}_{\vec{F}_r}(\vec{u}), \label{eq:radiativeHydrodynamicsEqua:radiativeFlux}
  \end{empheq}
  \label{eq:radiativeHydrodynamicsEqua}
\end{subequations}
where $\vec{u}$ is the material velocity, $p$ is the pressure of the fluid, $\vec{g}$ is the external gravitational field, $\rho E = \rho e + \frac{1}{2} \rho \vec{u}^2$ is the density of total matter energy with $e$ the specific internal energy. The terms $S_{E_r}(\vec{u})$ and $\vec{S}_{\vec{F}_r}(\vec{u})$ depend on the velocity $\vec{u}$. Using Eq. 29 and Eq. 31 from \citet{lowrie1999}, we have
\begin{equation}
  \left\{
    \begin{aligned}
      S_{E_r}(\vec{u}) &= \frac{\sigma}{c} \vec{u} \cdot \vec{F}_r + \frac{\sigma}{c} E_r \vec{u} \cdot \vec{u} + \frac{\sigma}{c} \vec{u} \cdot (\vec{u} \cdot \tens{P}_r) \\
      \vec{S}_{\vec{F}_r} (\vec{u}) &=  \sigma \vec{u} \cdot \tens{P}_r + \sigma a_r T_g^4 \vec{u} + \frac{\sigma}{c^2} \vec{u} \cdot (\vec{u} \cdot \vec{F}_r),
    \end{aligned}
  \right.
\end{equation}
at first order in $\frac{\vec{u}}{c}$. To close the system, we also add the equation of state of an ideal gas: $p = \rho e (\gamma-1)$.

\subsection{Diffusive limit in a static fluid} \label{sect:diffusiveLimit}

Let us now focus on the diffusive regime with the hydrodynamics frozen, i.e. the limit of large opacity, long timescale, and $\vec{u} = 0$. We consider the P\textsubscript{1} closure relation
\begin{equation}
  \left\{
    \begin{aligned}
      \partial_t E_r + \nabla \cdot \vec{F}_r &= c \sigma \left( a_r T_g^4 - E_r \right)\\
      \partial_t \vec{F}_r + \frac{c^2}{3} \nabla E_r &= -c \sigma \vec{F}_r\\
      \partial_t (\rho c_v T_g) &= - c \sigma \left( a_r T_g^4 - E_r \right).
    \end{aligned}
    \label{eq:p1Syst}
  \right.
\end{equation}
Following \citet{berthon2011}, we introduce a rescaling parameter $\varepsilon$ to write the time (resp. the opacity) as $\tilde{t} = \varepsilon t$ (resp. $\tilde{\sigma} = \varepsilon \sigma$). The radiative energy, the radiative flux, and the gas temperature are expanded with $\varepsilon$, e.g. $E_r = E_{r,0} + \varepsilon E_{r,1} + \mathcal{O}(\varepsilon^2)$. System \ref{eq:p1Syst} becomes
\begin{subequations}
  \begin{empheq}[left=\empheqlbrace]{align}
    \varepsilon^2 \partial_{\tilde{t}} E_r + \varepsilon \nabla \cdot \vec{F}_r &  = \begin{aligned}[t]
      & c \tilde{\sigma} \left( a_r T_g^4 - E_r \right) \label{eq:asymptoticDev:energy}\\
    \end{aligned}\\
    \varepsilon^2 \partial_{\tilde{t}} \vec{F}_r + \varepsilon \frac{c^2}{3} \nabla E_r &=  -c \tilde{\sigma} \vec{F}_r \label{eq:asymptoticDev:flux}\\
    \varepsilon^2 \partial_{\tilde{t}} (\rho c_v T_g) &=  -c \tilde{\sigma} \left( a_r T_g^4 - E_r \right). \label{eq:asymptoticDev:temperature}
  \end{empheq}
\end{subequations}
By expanding \autoref{eq:asymptoticDev:energy} and \autoref{eq:asymptoticDev:flux} at order $0$, we have
\begin{equation}
  \left\{
    \begin{aligned}
      E_{r,0} &=  a_r T_{g,0}^4\\
      \vec{F}_{r,0} &= \vec{0}.
    \end{aligned}
  \right.
  \label{eq:asymptoticDevOrder0}
\end{equation}
Expanding \autoref{eq:asymptoticDev:flux} at order $1$ leads to
\begin{equation}
  \vec{F}_{r,1} = -\frac{c}{3 \tilde{\sigma}} \nabla E_{r,0}.
  \label{eq:asymptoticDevOrder1}
\end{equation}
Finally, expanding the sum of \autoref{eq:asymptoticDev:energy} and \autoref{eq:asymptoticDev:temperature} at order $2$ gives
\begin{equation}
  \partial_{\tilde{t}} \left( E_{r,0} + \rho c_v T_{g,0} \right) - \nabla \left( \frac{c}{3 \tilde{\sigma}} \nabla E_{r,0} \right) = 0.
  \label{eq:diffusionEqua}
\end{equation}

In the diffusive limit, the total energy $E_r + \rho c_v T_g$ at order $0$ obeys the diffusion equation given by \autoref{eq:diffusionEqua}. A similar development for the radiation hydrodynamics case is done in \autoref{appendix:diffusiveLimitRHD}. The aim of the \autoref{sect:numericalScheme} is to design an asymptotic preserving scheme, i.e. a numerical scheme that will degenerate to the discretization of \autoref{eq:diffusionEqua} in the diffusive regime.

  \section{Numerical scheme and algorithm} \label{sect:numericalScheme}

  \subsection{Radiation transport in a static fluid} \label{sect:schemeRadiation}

  Let us first introduce some notations: we note $\Delta x$ the step along the x-direction. $\Delta t$ is the time interval between the current time $t^n$ and $t^{n+1}$. We write $x_i$ the center of the cell i and $x_{i+\frac{1}{2}}$ the interface between the cell $i$ and the cell $i+1$. We use the notation $u_i^n$ to represent the averaged quantity associated with the field $u$ at time $t^n$ in the cell $i$ (finite volume). Finally, we note $u_{i+\frac{1}{2}}^n$ to represent the quantity associated with the field $u$ at time $t^n$ and at the interface between cells $i$ and $i+1$. 

  The development of the numerical scheme is presented only in the one-dimensional case, but its extension to higher dimensions is straightforward. To ease notations, we drop the indices $r$ for all radiative variables.

  We present the development of the numerical scheme using a time-implicit integration, but a similar development can be done with a semi-implicit solver: source terms remain implicit, but the hyperbolic part is time-explicit.

  \subsubsection{Hyperbolic system} \label {sect:hyperbolicSystem}

  Following \citet{gonzalez2007}, we discretize the hyperbolic part of \autoref{eq:momentModel} using a first-order Godunov type solver \citep{toro2009}. From \citet{berthon2011}, we also introduce an extra parameter $\alpha$ which will be specified in \autoref{sect:ap}:
  \begin{equation}
    \left\{
      \begin{aligned}
        E_{i}^{n+1} &= E_{i}^n - \frac{\Delta t}{\Delta x} \left( \alpha_{i+\frac{1}{2}} \mathcal{F}^*_{i+\frac{1}{2}} - \alpha_{i-\frac{1}{2}} \mathcal{F}^*_{i-\frac{1}{2}} \right) \\
        &+ c \sigma_i \Delta t \left( a_r \left( T_{i}^{n+1} \right)^4 - E_{i}^{n+1} \right)\\
        F_{i}^{n+1} &= F_{i}^n - \frac{\Delta t}{\Delta x}  \left(\mathcal{P}^*_{i+\frac{1}{2}} - \mathcal{P}^*_{i-\frac{1}{2}} \right) - c \Delta t \{\sigma F\}_i^{n+1}\\
        \rho c_v T_{i}^{n+1} &= \rho c_v T_{i}^n    - c \sigma_i \Delta t \left( a_r \left( T_{i}^{n+1} \right)^4 - E_{i}^{n+1} \right),
      \end{aligned}
    \right.
  \end{equation}
  where $\mathcal{F}_{i+\frac{1}{2}}^*$ and $\mathcal{P}^*_{i+\frac{1}{2}}$ are the numerical fluxes given by
  \begin{equation}
    \begin{alignedat}{5}
      \mathcal{F}^*_{i+\frac{1}{2}} &=& \frac{\lambda^+_{i+\frac{1}{2}}F_{i}^{n+1} - \lambda^-_{i+\frac{1}{2}}F_{i+1}^{n+1}}{\lambda^+_{i+\frac{1}{2}} - \lambda^-_{i+\frac{1}{2}}} &+& \frac{\lambda^+_{i+\frac{1}{2}} \lambda^-_{i+\frac{1}{2}}}{\lambda^+_{i+\frac{1}{2}} - \lambda^-_{i+\frac{1}{2}}} \left( E_{i+1}^{n+1} - E_{i}^{n+1} \right)\\
      \mathcal{P}^*_{i+\frac{1}{2}} &=& c^2 \frac{\lambda^+_{i+\frac{1}{2}}P_{i}^{n+1} - \lambda^-_{i+\frac{1}{2}}P_{i+1}^{n+1}}{\lambda^+_{i+\frac{1}{2}} - \lambda^-_{i+\frac{1}{2}}} &+& \frac{\lambda^+_{i+\frac{1}{2}} \lambda^-_{i+\frac{1}{2}}}{\lambda^+_{i+\frac{1}{2}} - \lambda^-_{i+\frac{1}{2}}} \left( F_{i+1}^{n+1} - F_{i}^{n+1} \right).
    \end{alignedat}
  \end{equation}
  with $\lambda_{i+\frac{1}{2}}^+ = \max\left( 0, \lambda_{max} \right)$ and $\lambda_{i+\frac{1}{2}}^- = \min\left( 0, \lambda_{min} \right)$, where $\lambda_{max}$ and $\lambda_{min}$ are the eigenvalues of \autoref{eq:momentModel}. From \citet{berthon2007}, we have
  \begin{equation}
    \lambda_{max, min} = c\left( \frac{f_x}{\xi} \pm \frac{\sqrt{2} \sqrt{(\xi-1)(\xi+2)(2(\xi-1)(\xi+2)+3f_y^2)}}{\sqrt{3}\xi(\xi+2)} \right),
    \label{eq:eigenvalues}
  \end{equation}
  with $\xi = \sqrt{4-3f^2}$. See Fig. 1 of \citet{gonzalez2007} for more details about the structure of the eigenvalues. $\{\sigma F\}_i^{n+1}$ is a well-chosen discretization of the term $\sigma \vec{F}_r$ in the cell $i$ and at time $t^{n+1}$. It will be specified in the next section.

  \subsubsection{Well-balanced modification of the source term} \label{sect:wb}

  From \citet{berthon2015}, a well-balanced scheme catches the correct steady regime. The steady state, if it exists, is given by
  \begin{subequations}
    \begin{empheq}[left=\empheqlbrace]{align}
      E_r &  = \begin{aligned}[t]
        a_r T_g^4 \label{eq:steadyState:energy}\\
      \end{aligned}\\
      \nabla \cdot \vec{F}_r &= 0 \label{eq:steadyState:flux}\\
      c \nabla \cdot \tens{P}_r &= - \sigma \vec{F}_r \label{eq:steadyState:pressure}.
    \end{empheq}
    \label{eq:steadyState}
  \end{subequations}
  \Autoref{eq:steadyState:pressure} is discretized by $c \frac{P_{i+1}^{n+1} - P_{i-1}^{n+1}}{2 \Delta x} = - \{\sigma F\}_i^{n+1}$. An obvious choice for $\{\sigma F\}_i^n$ is
  \begin{equation}
    \{\sigma F\}_i^{n+1} = \sigma_i F_i^{n+1}.
    \label{eq:wb_naive}
  \end{equation}
  However, using this formulation, \autoref{eq:steadyState:pressure} is discretized as 
  \begin{equation}
    - \frac{c}{2} \left( \left( \nabla \cdot P \right)_{i+\frac{1}{2}}^{n+1} + \left( \nabla \cdot P \right)_{i-\frac{1}{2}}^{n+1} \right) = \sigma_i F_i,
  \end{equation}
with $\left(\nabla \cdot P\right)_{i+\frac{1}{2}}^{n+1} = \frac{P_{i+1}^{n+1} - P_i^{n+1}}{\Delta x}$. The radiative flux remains cell-centered and is equal to the divergence of radiative pressure, defined at the interfaces of the cells. This can create some spurious flux at the interface when looking for a steady state with a constant flux in the box (see \autoref{sect:wbTest}). Inspired by well-balanced schemes for hydrodynamics (e.g. \citealt{padioleau2019}) which preserve the hydrostatic balance between the pressure forces and the gravitational force (and the similarity of this balance with the balance between radiative pressure and radiative flux source term in \autoref{eq:steadyState:pressure}), we choose to use an average of a face discretization of the radiative flux source term:
  \begin{equation}
    \{\sigma F\}_{i}^{n+1} = \frac{1}{2} \left( \sigma_{i+\frac{1}{2}} F_{i+\frac{1}{2}}^{n+1} + \sigma_{i-\frac{1}{2}} F_{i-\frac{1}{2}}^{n+1} \right),
    \label{eq:wb_smart}
  \end{equation}
  with 
  \begin{equation}
    \left\{
      \begin{aligned}
        \sigma_{i+\frac{1}{2}} &= \frac{1}{2} \left( \sigma_i + \sigma_{i+1} \right)\\
        F_{i+\frac{1}{2}}^{n+1} &= \frac{1}{2} \left( F_i^{n+1} + F_{i+1}^{n+1} \right).\\
      \end{aligned}
    \right.
  \end{equation}
  One way to interpret this equation is to remember that 
    \begin{equation}
      \{\sigma F\}_i^{n+1} = \frac{1}{\Delta x} \int_{x_{i-\frac{1}{2}}}^{x_{i+\frac{1}{2}}} \sigma(x) \vec{F}_r\left(t^{n+1}, x\right) dx.
      \label{eq:wb_integration}
    \end{equation}
    \Autoref{eq:wb_naive} is obtained with the rectangle rule for numerical integration of \autoref{eq:wb_integration}:
    \begin{equation}
      \begin{aligned}
        \{\sigma F\}_i^{n+1} &= \frac{ x_{i+\frac{1}{2}} - x_{i-\frac{1}{2}}}{\Delta x} \left( \sigma\left( \frac{x_{i-\frac{1}{2}} + x_{i+\frac{1}{2}}}{2} \right) \vec{F}_r\left( t^{n+1}, \frac{x_{i-\frac{1}{2}} + x_{i+\frac{1}{2}}}{2} \right) \right) \\
        &= \sigma(x_i) \vec{F}_r(t^{n+1}, x_i)\\
        &= \sigma_i F_i^{n+1}.
      \end{aligned}
    \end{equation}
    whereas \autoref{eq:wb_smart} is given by the trapezoidal rule:
    \begin{equation}
      \begin{aligned}
        \{\sigma F\}_i^{n+1} &= \frac{ x_{i+\frac{1}{2}} - x_{i-\frac{1}{2}}}{2 \Delta x} \left( \sigma\left( x_{i-\frac{1}{2}} \right) \vec{F}_r \left( t^{n+1}, x_{i-\frac{1}{2}} \right) + \sigma\left( x_{i+\frac{1}{2}} \right) \vec{F}_r\left( t^{n+1}, x_{i+\frac{1}{2}} \right) \right)\\
        &= \frac{1}{2} \left( \sigma_{i-\frac{1}{2}} F_{i-\frac{1}{2}}^{n+1} + \sigma_{i+\frac{1}{2}} F_{i+\frac{1}{2}}^{n+1} \right).
      \end{aligned}
    \end{equation}

  To have 
  \begin{equation}
    \sigma_{i+\frac{1}{2}} F_{i+\frac{1}{2}}^{n+1} = -c \frac{ P_{i+1}^{n+1} - P_{i}^{n+1}}{\Delta x}
  \end{equation}
  in the whole domain, we also impose it as boundary condition:
  \begin{equation}
    \sigma_{\frac{1}{2}} F_{\frac{1}{2}}^{n+1} = -c  \frac{P_1^{n+1} - P_0^{n+1}}{\Delta x},
    \label{eq:boundaryCondition}
  \end{equation}
  where $P_0^{n+1}$ is the radiative pressure given by the boundary condition. In that way, the radiative flux is centered at the interfaces of the cells, as well as the divergence of radiative pressure. A von Neumann stability analysis of the modified scheme is done in \autoref{appendix:wbStability}.

  \subsubsection{Asymptotic preserving scheme} \label{sect:ap}

  We still have to specify our choice for $\alpha_{i+\frac{1}{2}}$. $\alpha_{i+\frac{1}{2}}=1$ corresponds to a classic HLL scheme. However, the solution given by an asymptotic preserving scheme has to approximate the solution of \autoref{eq:diffusionEqua} as soon as the asymptotic regime is reached, i.e. large opacity and long timescale. Unfortunately, a standard HLL scheme does not have this property (see \autoref{sect:marshakWave}). To tackle this issue and get an asymptotic preserving scheme, we choose
  \begin{equation}
    \alpha_{i+\frac{1}{2}} = \cfrac{1}{1- 3 \sigma_{i+\frac{1}{2}} \Delta x \left( 1-f_{i+\frac{1}{2}}^2 \right) \cfrac{\lambda_{i+\frac{1}{2}}^+ \lambda_{i+\frac{1}{2}}^-}{c \left(\lambda_{i+\frac{1}{2}}^+ - \lambda_{i+\frac{1}{2}}^-\right)}}.
    \label{eq:asymptoticCorrection}
  \end{equation}
  with $f_{i+\frac{1}{2}} = \frac{1}{2}\left( f_i^n + f_{i+1}^n \right)$. The derivation of \autoref{eq:asymptoticCorrection} is done in \autoref{appendix:proofAP}. Other choices can be done (see for example \citealt{berthon2011}). If $\sigma_{i+\frac{1}{2}} \Delta x$ goes to $0$, $\alpha_{i+\frac{1}{2}}$ goes to $1$, and we recover a standard HLL scheme. Considering the diffusive limit, we prove that the scheme is asymptotic preserving in \autoref{appendix:proofAP}. We show that
  \begin{equation}
    \left\{
      \begin{aligned}
        E_{i,0}^{n+1} &= a_r \left( T_{i,0}^{n+1} \right)^4\\
        F_{i,0}^{n+1} &= 0\\
        \sigma_{i+\frac{1}{2}} F_{i+\frac{1}{2}, 1}^{n+1} &= - \frac{c}{3 \Delta x} \left( E_{i+1,0}^{n+1} - E_{i,0}^{n+1} \right)\\
        E_{i,0}^{n+1} + \rho c_v T_{i,0}^{n+1} &= E_{i,0}^n + \rho c_v T_{i,0}^n\\
        &+ \frac{c \Delta t}{3 \Delta x^2} \left( \frac{E_{i+1,0}^{n+1} - E_{i,0}^{n+1}}{\sigma_{i+\frac{1}{2}}} - \frac{E_{i,0}^{n+1} - E_{i-1,0}^{n+1}}{\sigma_{i-\frac{1}{2}}} \right),
      \end{aligned}
    \right.
    \label{eq:discreteDiffusionEqua}
  \end{equation}
  which is a standard discretization of \autoref{eq:asymptoticDevOrder0}, \autoref{eq:asymptoticDevOrder1} and \autoref{eq:diffusionEqua}.

  Unfortunately, we cannot prove that this scheme will preserve the admissible states $f < 1$ and, indeed, numerical experiments with this scheme have shown that we can get $f > 1$ when we are close to the free-streaming regime. In these situations, we can either enforce $f <1$ (\autoref{sect:hiiRegionExpansion}) or come back to a centered discretization of the source term (\autoref{sect:shadow}). Furthermore, the development of the asymptotic preserving scheme with the well-balanced modification of the source term is only done in the case of a static fluid. Using the asymptotic correction \autoref{eq:asymptoticCorrection} is only the first step to have an asymptotic preserving scheme in the case of a moving fluid (see \autoref{sect:apVelocity}).

  \subsection{Coupling to hydrodynamics} \label{sect:couplingHydro}

  Following \citet{gonzalez2007}, the resolution of the whole system \ref{eq:radiativeHydrodynamicsEqua} describing radiation hydrodynamics is split into three steps:
  \begin{enumerate}
    \item update of the hydrodynamics quantities (\autoref{eq:radiativeHydrodynamicsEqua:mass}, \autoref{eq:radiativeHydrodynamicsEqua:momentum} and \autoref{eq:radiativeHydrodynamicsEqua:energy} without the terms of energy and momentum exchange) using the well-balanced and all-regime solver developed in \citet{padioleau2019};
    \item update of the radiative quantities and gas temperature (\autoref{eq:momentModel} and \autoref{eq:internalEnergyEqua}) using the solver developed in \autoref{sect:schemeRadiation}. During this step, the hydrodynamics quantities are frozen;
    \item addition of source terms $S_{E_r}(\vec{u})$ and $\vec{S}_{\vec{F}_r}(\vec{u})$. For simplicity, all source terms which depend on the velocity are treated explicitly. The term $\frac{\sigma}{c} \vec{F}_r$ in \autoref{eq:radiativeHydrodynamicsEqua:momentum} and \autoref{eq:radiativeHydrodynamicsEqua:radiativeFlux} is discretized using the well-balanced scheme proposed in \autoref{sect:wb}. All the other terms remain cell-centered.
  \end{enumerate}
  This splitting allows reducing the number of equations solved implicitly, making the method more efficient.

  \subsection{Algorithm for non-linear implicit solver} \label{sect:implicitSolver}

  The time step given by the CFL condition is much smaller for radiation than for hydrodynamics. Indeed, for the radiation, it is limited by the speed of light, whereas it is limited by the speed of sound of the fluid for the hydrodynamics. Because we are interested in radiation hydrodynamics, we will consider a long timescale for the radiative transfer. Therefore, we use a time-implicit integration {for the radiative transfer. 

  \subsubsection{Newton-Raphson method and linear solver} \label{sect:newton}

  Because of the Eddington tensor, the eigenvalues in the numerical fluxes, and the $a_r T_g^4$ factor, the system is non-linear. It is solved using a Newton-Raphson method. At each iteration, we have to solve a linear system. Because the system is large ($(2+d)N$ unknowns, where $d$ is the number of dimensions and $N$ the total number of cells) and sparse, it cannot be solved using a direct method. Because of the numerical fluxes, the matrix is not symmetric, and we use the biconjugate gradient stabilized method \citep{vanDerVorst1992}.

  \subsubsection{Preconditioner} \label{sect:preconditioner}

  Using large time steps for the radiative transfer, the matrix is ill-conditioned and iterative methods might not converge. One way to deal with this issue is to use a preconditioner. Instead of solving the original linear system $\tens{A} \vec{x} = \vec{b}$, we solve the right preconditioned system $\tens{A K}^{-1} \tens{K} \vec{x} = \vec{b}$ via solving $\tens{A K}^{-1} \vec{y} = \vec{b}$ to compute $\vec{y}$ and then $\tens{K} \vec{x} = \vec{y}$. As long as the matrix $\tens{K}$ is invertible, this gives the same solution as the original system. If $\tens{K}$ is well-chosen, the condition number of the matrix $\tens{AK}^{-1}$ is lower than $\tens{A}$'s one. For example, the Jacobi or diagonal preconditioner is given by
  \begin{equation}
    \tens{K}_{ij} = 
    \left\{
      \begin{aligned}
        \tens{A}_{ij} &\ \mathrm{ if}\ i = j\\
        0 &\ \mathrm{ otherwise}.
      \end{aligned}
    \right.
  \end{equation}
  See \citet{saad2003} for more details.

  Numerous preconditioners have been developed over the years, some of them perform well, some do not, depending on the problem considered. Among different preconditioners, we will explore the algebraic multigrid (AMG) technique. We present this method in the next section, and we compare it with other preconditioners in \autoref{sect:performances}: a standard ILU(k) factorization \citep{saad2003}, a slightly modified variant of the standard ILU factorization \citep{saad1994}, an additive Schwarz domain decomposition \citep{cai1999}, and a classical relaxation method, Jacobi with damping \citep{saad2003}.

  \subsubsection{Algebraic multigrid (AMG) preconditioner} \label{sect:amgPrec}

  The algebraic multigrid (AMG) methods were first developed as linear solvers for symmetric positive definite matrices arising from the discretization of scalar elliptic PDEs. For such a matrix, classical iterative methods are efficient to compute the high frequencies of the solution, but lack efficiency to compute its low frequencies. However, the computation is easier on a coarser grid with fewer unknowns. The idea of the multigrid solver is to build a coarser grid, then solve the problem on this coarse grid and finally interpolate the solution on the fine grid. We can then define a restriction operator $\tens{R}$ which transfers vectors from the fine grid to the coarse grid and an interpolation operator $\tens{P}$ used to return to the finer grid. $\tens{P}$ and $\tens{R}$ are non squared matrices. From \citet{saad2003}, here are the main steps of the method:
  \begin{enumerate}
    \item pre-smoothing: a few iterations of a simple method such as Jacobi or an incomplete factorization are performed, to get the value $\vec{\tilde{x}}$;
    \item the residual $\vec{\tilde{r}} = \vec{b} - \tens{A} \vec{\tilde{x}}$ is projected over the coarse grid with the restriction operator $\tens{R}$, to get the residual equation $\tens{RAP} \vec{y} = \tens{R} \vec{\tilde{r}}$;
    \item this equation is solved, possibly with a direct solver;
    \item the solution $\vec{y}$ is interpolated over the fine grid with the interpolation operator $\tens{P}$ and then $\vec{\bar{x}} = \vec{\tilde{x}} + \tens{P} \vec{y}$;
    \item post-smoothing: a few iterations of a simple method are again performed to get the solution $\vec{\tilde{\bar{x}}}$.
  \end{enumerate}
  The solution $\vec{\tilde{\bar{x}}}$ is used as a preconditioner result. If the coarse grid has too many unknowns to be solved directly, this process is applied recursively: the coarse grid becomes the fine grid and a coarser grid is built. Therefore, we have a hierarchy of grids. With a geometric multigrid solver, the restriction and interpolation operators are determined by the mesh, whereas, with an algebraic multigrid solver, they are automatically generated, using data from the matrix. 

  Before showing numerical tests, we present some details about implementation and parallelization in the next section.

  \section{Implementation and parallelization} \label{sect:implementationAndParallelization}

  Our implementation has been done in the code ARK-RT\footnote{\url{https://gitlab.erc-atmo.eu/erc-atmo/ark-rt/tree/v1.0.0}}, a fork of the code ARK developed in \citet{padioleau2019}. The hydrodynamics and gravity part of our solver is similar to ARK and is solved with a well-balanced and all-regime solver. Because the solver for the radiative transfer equations is time-implicit, we have to solve large sparse linear systems. It is done using the library Trilinos, described in the next section.

  \subsection{Linear algebra} \label{sect:linearAlgebra}

  We use the second generation of packages of the framework Trilinos \citep{heroux2005}, i.e. Kokkos \citep{edwards2014} for shared memory computation, Tpetra \citep{baker2012} for distributed vectors and matrices, Belos \citep{bavier2012} for linear solvers, Ifpack2 \citep{prokopenko2016} for classical preconditioners and MueLu \citep{MueLuURL, MueLu} for the AMG preconditioner. Let us focus on the package Kokkos.

  \subsubsection{Shared memory computation} \label{sect:kokkos}

  As new architectures have more and more cores, the distributed memory model is not enough to take advantage of all the computational power available. Therefore, we need to use a shared memory model inside the nodes. Furthermore, computational nodes are more and more heterogeneous, for example, multi-cores, many-cores, or accelerators such as GP-GPUs. Each architecture requires its own interface, such as OpenMP or C++11 threads for multi-cores and many-cores processors and CUDA or OpenACC for NVIDIA GPUs. This raises the problem of portability and performance portability: many HPC codes are optimized for some specific architectures, so running the code on a different architecture will result in bad performance.

  The library Kokkos \citep{edwards2014} tackles this issue. The user has a unique code which can be compiled with different shared memory models such as OpenMP or CUDA. Kokkos is based on different abstractions, like execution spaces (where a function is executed), memory spaces (where the data are), and execution policies (how the function is executed). It provides execution patterns such as parallel loops and multidimensional arrays, the storage is optimized according to the architecture.

  \subsection{Implementation} \label{sect:implementation}

  For the hydrodynamics step, Kokkos is used as an independent library for shared memory computation. Communications between the nodes are handled by the Message Passing Interface (MPI) programming model through a regular domain decomposition. Following \citet{kestener2017}, inside each node, the domains are endowed with ghost cells used to implement physical boundary conditions, but also to contain values from neighbor domains. The code is organized with computational kernels, each kernel is a C++ functor. See \citet{padioleau2019} for more details.

  The second step is the time-implicit solver for radiative transfer. The values of the matrix and the right-hand side of the linear system have to be updated at each iteration of the Newton-Raphson method, but the graph of the matrix does not change. Using Trilinos, the rows of the matrix are distributed across the MPI processes, each of them is associated with a unique global index. Each global index has a matching local index on the owning process.

  Trilinos provides several methods to update the coefficients of the matrix. One of them uses only global indices, which is the way recommended by Trilinos. However, this function can only be called by the host. This has two main consequences. First, because all the rows of the matrix have to be updated, we use a sequential loop over the rows of the matrix. Second, in the case where we are using GP-GPUs, we have to update the matrix with data coming from the device. Therefore, we have to transfer some data from the device to the host, update the matrix with these data, and then transfer the matrix from the host to the device, this last step is done implicitly by Trilinos. This will increase the computational cost.

  Another way is to use local indices. The package KokkosKernels, part of the Kokkos ecosystem, provides several ways to update the coefficients of the matrix through a kernel. This allows the use of a parallel loop (via Kokkos::parallel\_for) and we avoid data transfers between the host and the device. For performance reasons, we use local indices.

  \subsection{Performances} \label{sect:performances}

  Thanks to Trilinos, we can use many preconditioners. Unfortunately, they do not behave the same way when the size of the system increases. All tests are performed on Poincare, our local cluster at Maison de la Simulation. Each node consists of two Sandy Bridge E5-2670 @ $2.60\ \mathrm{GHz}$ ($2 \times 8$ cores, 32 Go RAM) processors. We use a hybrid configuration MPI/OpenMP, with one MPI process per socket to avoid NUMA effects. 

  \begin{figure*}
    \centering
    \includegraphics{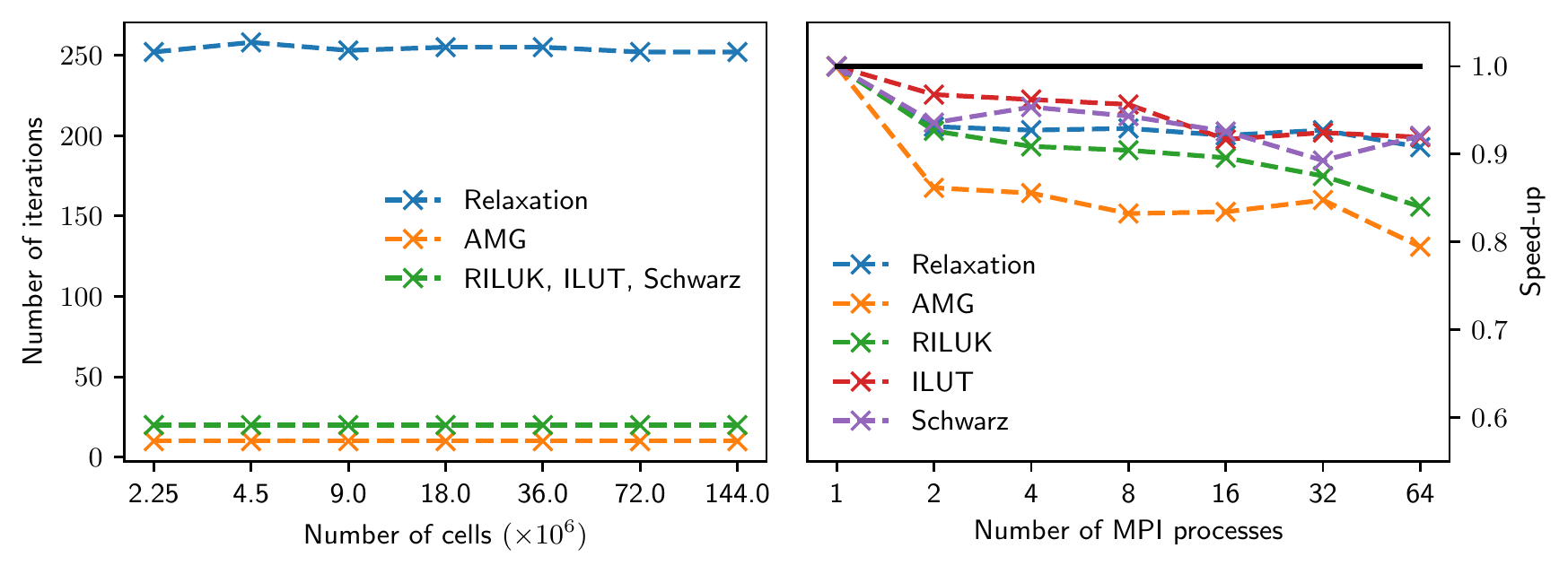}
    \caption{Weak scaling test. Each MPI process treats $1500^2$ cells. We have tested different preconditioners: Jacobi with damping (Relaxation), algebraic multigrid (AMG), standard ILU(k) factorization (RILUK), variant of the standard ILU factorization (ILUT) and additive Schwarz domain decomposition (Schwarz).
      Left panel: number of iterations to solve the linear system as a function of the number of cells.
    Right panel: speed-up as a function of the number of MPI processes.}
    \label{fig:weak_scaling}
  \end{figure*}

  We first performed a weak scaling test, where we consider a two-dimensional case with periodic boundary conditions and a hot source located at the center of each domain. Each MPI process is getting a piece of the whole domain of $1500^2$ cells, therefore the size of the system increases with the number of MPI tasks. The resolution is close to the one we are aiming for three-dimensional simulations. The left panel of \autoref{fig:weak_scaling} shows the mean number of iterations for the linear solver to converge as a function of the number of cells. For all preconditioners, the number of iterations remains constant, around $10$ iterations for the AMG preconditioner, around $20$ iterations for both incomplete factorizations and the additive Schwarz domain decomposition and around $250$ iterations for the relaxation. The right panel of \autoref{fig:weak_scaling} shows the speed-up as a function of the number of MPI processes. The speed-up reaches a plateau of $80\%$ to $90\%$ of maximum performance, depending on the preconditioner.

  \begin{figure*}
    \centering
    \includegraphics{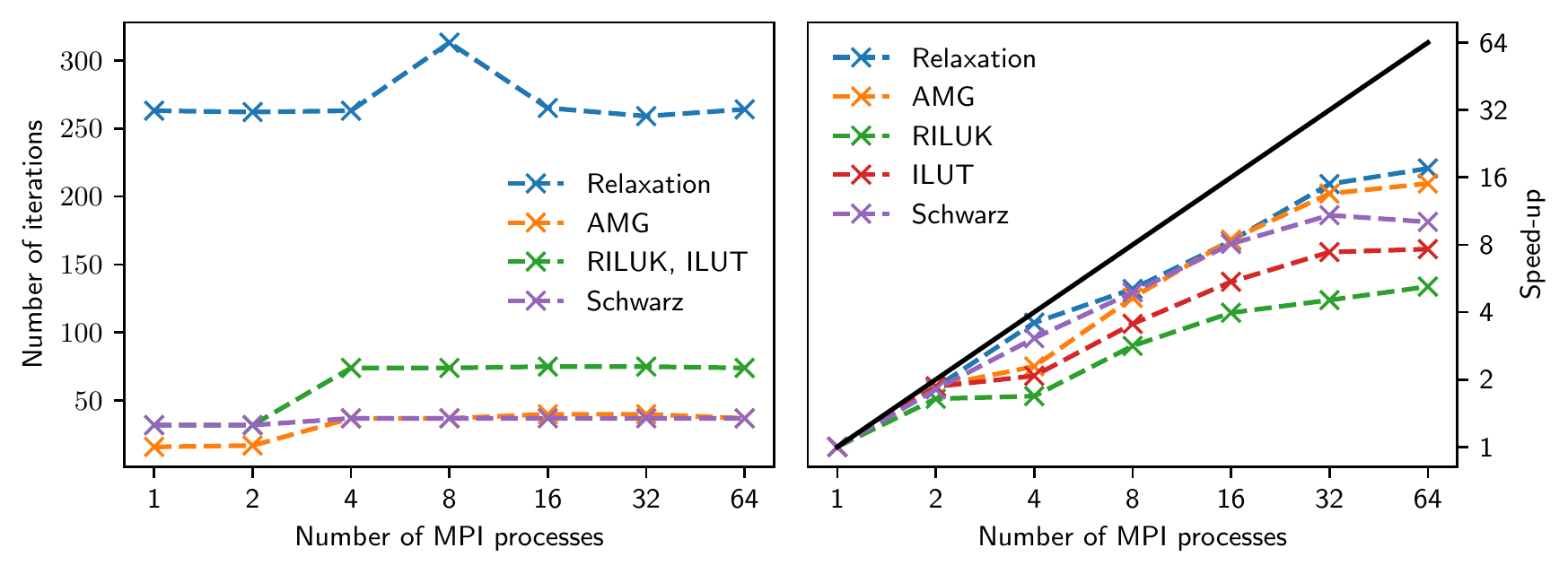}
    \caption{Strong scaling test. The global resolution is $2048^2$ cells. We have tested different preconditioners: Jacobi with damping (Relaxation), algebraic multigrid (AMG), standard ILU(k) factorization (RILUK), variant of the standard ILU factorization (ILUT) and additive Schwarz domain decomposition (Schwarz).
      Left panel: number of iterations to solve the linear system as a function of the number of MPI processes.
    Right panel: speed-up as a function of the number of MPI processes.}
    \label{fig:strong_scaling}
  \end{figure*}

  \Autoref{fig:strong_scaling} shows the number of iterations and the speed-up as a function of the number of MPI processes for a strong scaling test. We now consider a Marshak wave propagation in the diffusive limit. The global resolution remains constant as the number of processes increases. It is set to $2048^2$ cells. Because the global resolution is constant, one can expect the number of iterations to also remain constant when the number of MPI processes increases. However, using the algebraic multigrid (orange curve) and the incomplete factorizations (green curve), when four MPI processes or more are used, the number of iterations is twice the number of iterations reached with one or two MPI processes. Therefore, the computational time is the same using two or four MPI processes. Furthermore, all tested preconditioners and the linear solver requires several communications per iteration, which likely become the main cost when the local resolution decreases.

  \begin{table*}
    \begin{tabular}{cccc}     
      \hline
      \noalign{\smallskip}
      Scheme      &  Number of time steps & Computational time OpenMP (s) & Computational time CUDA (s) \\
      \noalign{\smallskip}
      \hline
      \noalign{\smallskip}
      Explicit & $73823$ & $6991$ & $839$ \\
      Implicit (non-parallel update) & $1$ & $60$ & $93$ \\
      Implicit (parallel update) & $1$ & $44$ & $77$ \\
      \noalign{\smallskip}
    \end{tabular}
    \caption{Computational time with both explicit and implicit solvers on CPU and GPU. With the implicit solver, the matrix is updated in a parallel or a non-parallel way. The implicit solver uses the AMG preconditioner.}
    \label{table:perfUpdate}
  \end{table*}

  Thanks to Kokkos, we can use exactly the same code on different architectures like Sandy Bridge processors and NVIDIA GP-GPUs (e.g. K80). Unfortunately, the memory required by the AMG preconditioner with a $1500^2$ simulation is larger than the memory available on a K80 GPU. For the next tests, we use a lower resolution of $1000^2$ cells. \autoref{table:perfUpdate} summarizes the computational time for a fixed problem with different schemes and different architectures. As the explicit solver is restricted by a CFL condition, it needs several thousands of time steps whereas the implicit solver only needs a few time steps to reach the same final time. Updating the matrix in parallel allows for a $25\%$ reduction in computational time required. On CPU, the implicit solver is around $160$ times faster than the explicit solver, whereas, on GPU, it is only $11$ times faster.

  \begin{figure}
    \centering
    \includegraphics{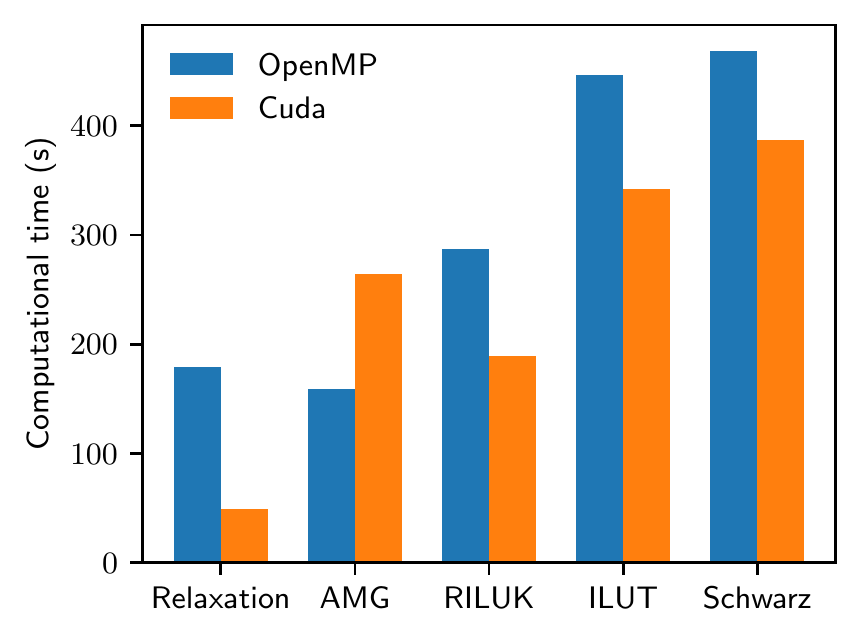}
    \caption{Computational time for the implicit solver with different preconditioners (Jacobi with damping (Relaxation),  algebraic multigrid (AMG), ILU(k) factorization (RILUK), slightly modified variant of ILU factorization (ILUT) and additive Schwarz domain decomposition (Schwarz)) on different architectures (Sandy Bridge CPU and K80 NVIDIA GPU). The resolution is $1000^2$ cells.}
    \label{fig:precond_arch_time}
  \end{figure}

  \Autoref{fig:precond_arch_time} compares the computational time with different preconditioners, on both CPU and GPU. Except for the implicit solver using the AMG preconditioner, all solvers are faster on GPU than CPU, up to three times faster for the relaxation preconditioner. Part of the AMG algorithm probably remains sequential. On CPU, the AMG preconditioner is faster than the relaxation preconditioner. The other preconditioners are slower, up to a factor $8$ between the relaxation and the additive Schwarz domain decomposition on GPU.

  \begin{figure}
    \centering
    \includegraphics{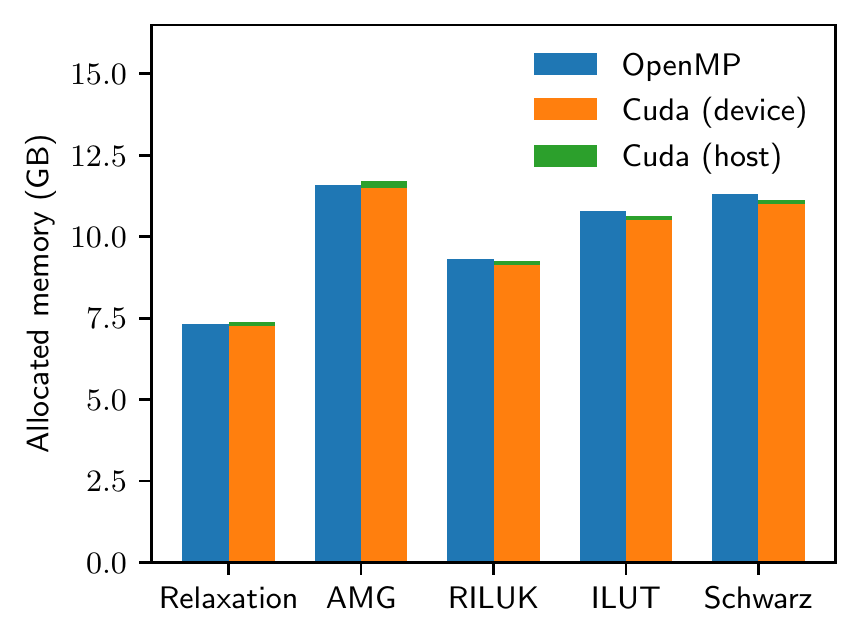}
  \caption{Memory consumption for the implicit solver with different preconditioners (Jacobi with damping (Relaxation),  algebraic multigrid (AMG), ILU(k) factorization (RILUK), slightly modified variant of ILU factorization (ILUT) and additive Schwarz domain decomposition (Schwarz)) on different architectures (Sandy Bridge CPU and K80 NVIDIA GPU). The resolution is $1000^2$ cells.}
    \label{fig:precond_arch_memory}
  \end{figure}

  \Autoref{fig:precond_arch_memory} compares the memory consumption with different preconditioners. Using GP-GPU, most of the data are located on the device, but Trilinos still allocates some memory on the host, between $0.125\ \mathrm{GB}$ and $0.208\ \mathrm{GB}$, unlike the explicit solver. Using the relaxation as a preconditioner, the amount of memory allocated is lower than with the other preconditioners ($7.3\ \mathrm{GB}$ for the relaxation against $11.5 \ \mathrm{GB}$ for the AMG).

  Choosing a well-suited preconditioner can be challenging and problem dependent. Once the preconditioner is chosen, it depends on many parameters. For example, Trilinos allows the user to choose the damping factor $\omega$ for the relaxation method or the smoother and the coarse solver for the AMG. Performances and stability can largely depend on these choices. For example, the relaxation method seems to be well suited for this problem with low computational time, and memory consumption, but in many other test cases, the linear solver will not converge. The AMG preconditioner performs well on CPU but is less efficient on GPU. Both incomplete factorizations and the additive Schwarz domain decomposition are slightly less efficient than the AMG preconditioner. Overall we have found the AMG preconditioner or relaxation method are a good compromise between stability and performances.

  The performances we obtained thanks to Kokkos and Trilinos are encouraging for the study of astrophysical problems. The time step given by the hydrodynamics can be written as $CFL \frac{\Delta x}{c}$. Using a relaxation as a preconditioner, we need $CFL \ge 50$ on CPU and $CFL \ge 100$ on GPU to save computational time, whereas, using an incomplete factorization, we need $CFL \ge 250$ on CPU and $CFL \ge 1000$ on GPU. We need a larger CFL number on GPU because the explicit solver is more efficient on GPU than CPU. 
  
  In the next section, we use several numerical tests to show that the scheme developed in \autoref{sect:numericalScheme} is well suited for the study of radiation hydrodynamics problems.

  \section{Numerical results} \label{sect:numericalResults}

  We performed a series of verification tests to validate different properties of the scheme: the asymptotic correction with a Marshak wave, the well-balanced property to reach a steady state with a jump of opacity, the properties of the M\textsubscript{1} model with a beam test and a shadow test, and the coupling to the hydrodynamics with radiative shocks. We also present a physical application about the stability of the ionization front in \ion{H}{ii} regions dense cores. To ease notations, we define the radiative temperature as $T_r = \left( \frac{E_r}{a_r} \right)^{\frac{1}{4}}$.

  \subsection{Marshak wave} \label{sect:marshakWave}

  From \citet{mihalas1984}, a Marshak wave is the propagation of hot radiation into a cold medium. We consider a one-dimensional case in the diffusive limit, to test the asymptotic preserving scheme developed in \autoref{sect:ap}.

  The length of the computational domain is $1\ \mathrm{cm}$, it is discretized with $400$ points. Initially, the medium is at equilibrium with the radiation: $T_0 = T_r = 300\ \mathrm{K}$, the initial radiative flux is $\vec{F}_r = \vec{0}$. We consider a perfect gas with $\gamma = \frac{5}{3}$. The hydrodynamics is frozen. The density is constant, such that $\rho c_v = 1\ \mathrm{J K^{-1} cm^{-3}}$, the opacity is also constant, with $\sigma = 10\,000\ \mathrm{cm^{-1}}$, therefore $\sigma \Delta x = 25$.
  At time $t = 0$, a source is lighted at the left boundary with $T_r = 1\,000\ \mathrm{K}$.

  \begin{figure}
    \centering
    \includegraphics{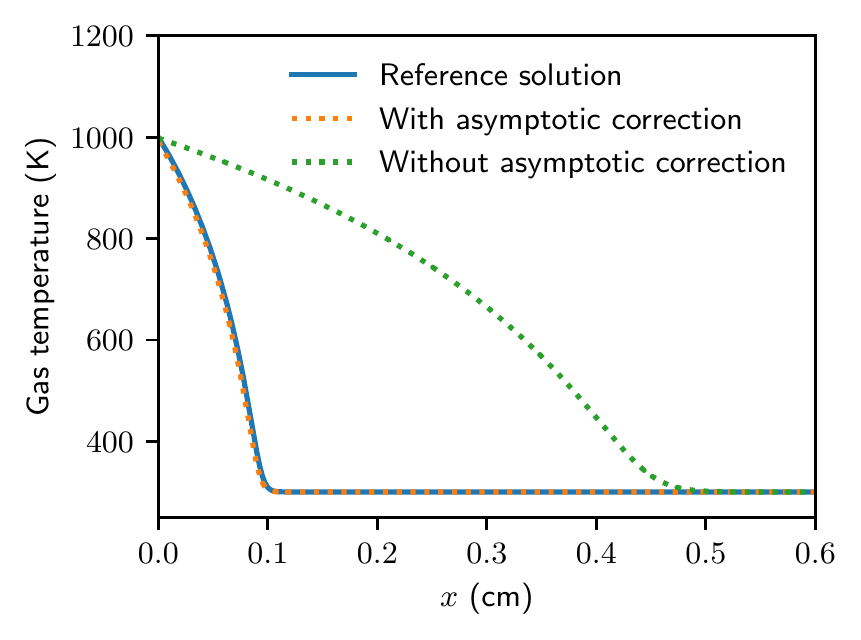}
    \caption{Marshak wave simulation. This figure shows a snapshot of the gas temperature at time $t_f = 2 \times 10^{-4}\ \mathrm{s}$, with and without the asymptotic correction and the reference solution. Spatial resolution is $n=400$ and the opacity is $\sigma=10\,000\ \mathrm{cm^{-1}}$.}
    \label{fig:marshakWave1d}
  \end{figure}

  The results are shown in \autoref{fig:marshakWave1d} at time $t_f = 2 \times 10^{-4}\ \mathrm{s}$. We compare different solutions: a reference solution, the solution given by our asymptotic preserving scheme, and the solution given by a standard scheme. The reference solution is given by a standard discretization of \autoref{eq:diffusionEqua}. From \citet{audit2002}, when a scheme that is not asymptotic preserving is used with $\sigma \Delta x \gg 1$, the grid does not sample the mean free path of the photons and the solution is dominated by numerical diffusion. The relative $L^2$ error between the reference solution and the solution with $\alpha_{i+\frac{1}{2}}$ given by \autoref{eq:asymptoticCorrection} is $1.1\%$ whereas, with the standard HLL scheme, the relative $L^2$ error is $84\%$. Using the asymptotic correction, we recover the correct behavior in the asymptotic limit.

  \subsection{Steady state with a jump of opacity} \label{sect:wbTest}

  In the previous case, the opacity is constant, we now consider a test with a jump of opacity, still in the one-dimensional case. We use this test to highlight the need for the well-balanced modification of the source term.

  The length of the computational domain is $1\ \mathrm{cm}$, it is discretized with $100$ points. Initially, the medium is at equilibrium with the radiation: $T_0 = T_r = 300\ \mathrm{K}$, the initial radiative flux is $\vec{F}_r = \vec{0}$. The opacity $\sigma$ is now a function of space:
  \begin{equation}
    \sigma(x) = 
    \left\{
      \begin{aligned}
        10\,000\ \mathrm{cm^{-1}} &\ \mathrm{ if} & x < 0.5,\\
        0 &\ \mathrm{ if} & x > 0.5.
      \end{aligned}
    \right.
  \end{equation}
  At time $t=0$, a source is lighted at the left boundary with $T_r = 1\,000\ \mathrm{K}$.

  \begin{figure}
    \centering
    \includegraphics{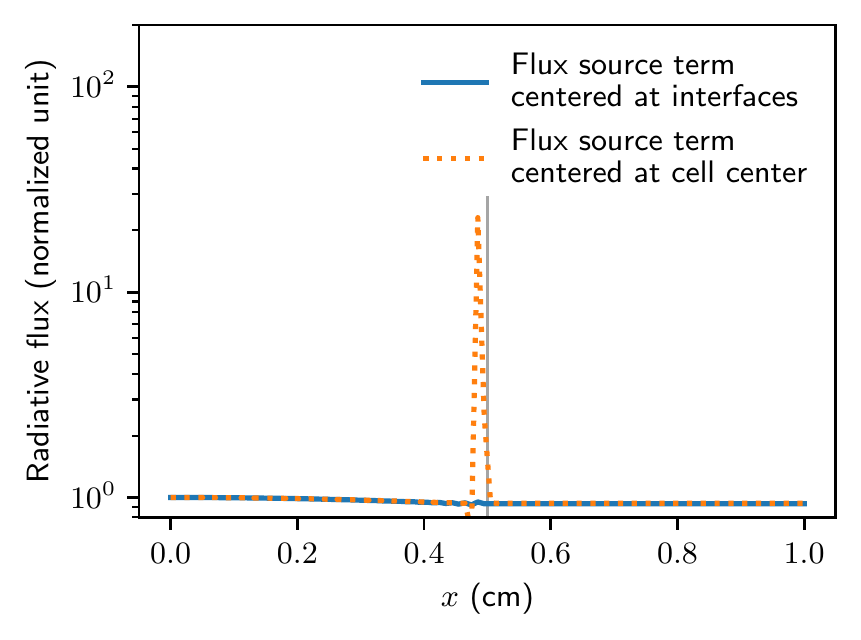}
    \caption{Simulation of a steady state with a jump of opacity. The opacity is piecewise constant, a jump is located at $x=0.5\ \mathrm{cm}$ (gray line). This figure shows a snapshot of the radiative flux at time $t_f = 10^{-3}\ \mathrm{s}$.}
    \label{fig:wellBalanced}
  \end{figure}

  \Autoref{fig:wellBalanced} shows the radiative flux at time $t_f = 10^{-3}\ \mathrm{s}$. From \autoref{eq:steadyState:flux}, when the steady state is reached, we expect the radiative flux to be constant in the box. Using a standard discretization of the source term, such as \autoref{eq:wb_naive}, a spurious peak located at the discontinuity of opacity is observed (orange curve). The value taken by the radiative flux is more than $20$ times the expected value. This seems to be caused by a numerical instability. This can result in $f > 1$ during the iterations of our Newton-Raphson implicit scheme, which is not physically admissible. However, using the well-balanced modification of the source term proposed by \autoref{eq:wb_smart} (blue curve), the spurious peak does not appear anymore and the constant steady state is reached.

  Using the standard discretization of the source term \autoref{eq:wb_naive}, one can show that the numerical scheme is unconditionally stable, in that the error between the numerical solution and the exact solution goes to $0$ as $\Delta x$ and $\Delta t$ go to $0$. The spurious peak seems to be due to a lack of precision in the integration of the source term. Using \autoref{eq:wb_smart}, the source term is defined at the interfaces of the cells and balances the divergence of radiative pressure, also defined at the interfaces.

  \subsection{Beam} \label{sect:beam}

  We now perform the same two-dimensional test as in \citet{richling2001, gonzalez2007}. The domain $[-1, 1] \times [-1, 1]$ is discretized with $128 \times 128$ cells. The initial temperature is $T_0 = T_r = 300\ \mathrm{K}$, the initial radiative flux is $\vec{F}_r = \vec{0}$, the opacity is $\sigma = 0$. At time $t=0$, a beam with $T = T_r = 1\,000\ \mathrm{K}$ is introduced with an angle of $45^\circ$. The beam is located at $x=-1$ and $y \in [-0.875, -0.75]$. Because we are in the free-streaming regime, we cannot use large time steps. For performance reasons, we use the semi-implicit scheme. 

  \begin{figure}
    \centering
    \includegraphics{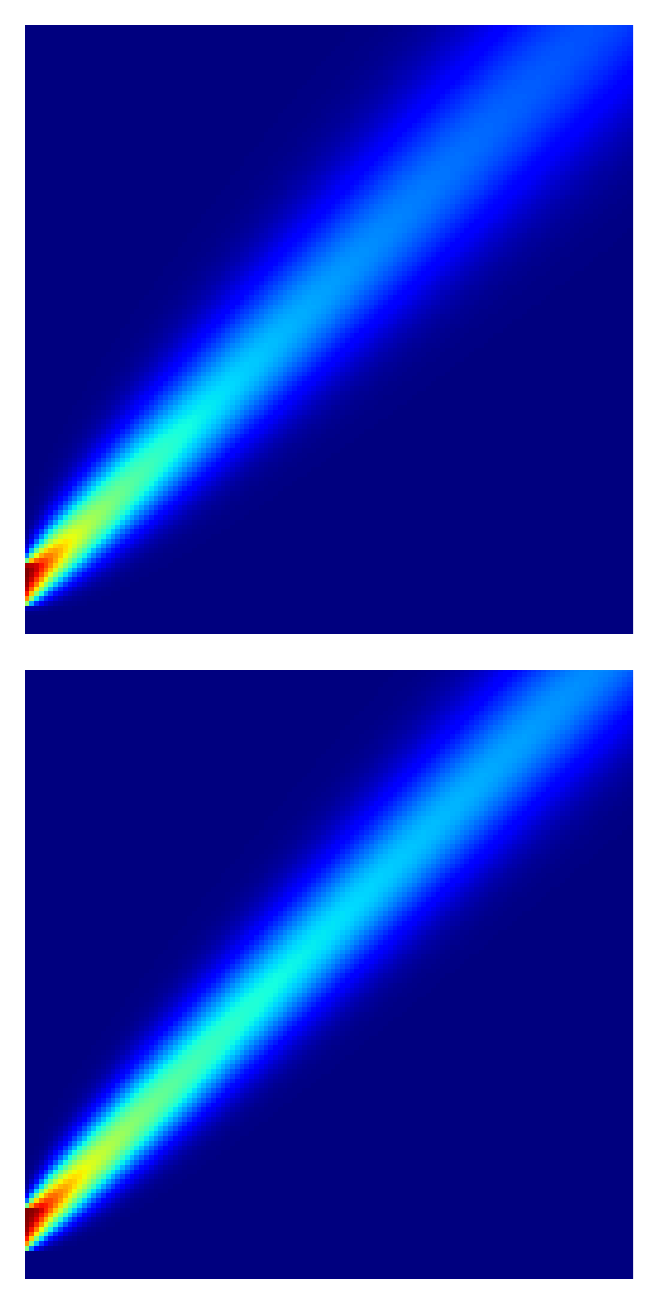}
    \caption{Beam simulation. The figure shows the radiative energy. The eigenvalues are fixed to $\pm c$ (upper panel) or calculated with \autoref{eq:eigenvalues} (lower panel).}
    \label{fig:beam}
  \end{figure}

  \begin{figure}
    \centering
    \includegraphics{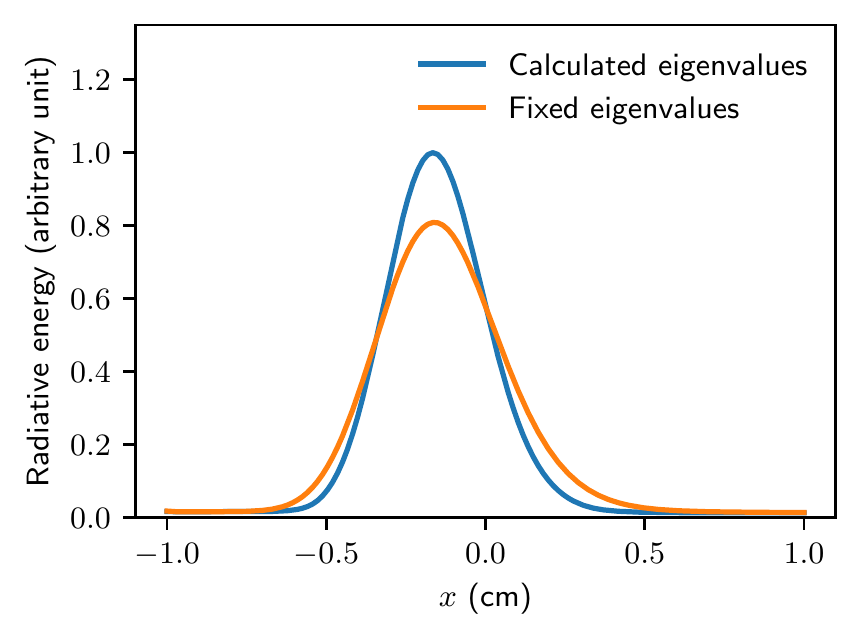}
    \caption{Beam simulation. The figure shows a horizontal cut in \autoref{fig:beam} at the middle height.}
    \label{fig:beamCut}
  \end{figure}

  Because there is no opacity, the beam propagates in the vacuum, and we expect it to cross the box without dispersion. The direction of the beam is not along the mesh axis, we use this test to quantify the numerical diffusion. \Autoref{fig:beam} shows the radiative energy at steady state, with the eigenvalues fixed to $\pm c$ and with computed eigenvalues. Because there is no opacity, the asymptotic correction nor the well-balanced source term affect the result, and we recover the same result as in \citet{gonzalez2007}. \Autoref{fig:beamCut} shows the horizontal cut at the middle height. The beam introduced at the boundary is sampled over $8$ cells. Ideally, without any numerical diffusion, we would expect the width of the beam to stay exactly $8$ cells. With the computed eigenvalues, we can keep the numerical diffusion under control. Using fixed eigenvalues, the width of the beam at middle height is approximately $30$ cells, whereas it is only $24$ cells with calculated eigenvalues.
  
  The main difference in this test between ARK-RT and HERACLES \citep{gonzalez2007} is the computation of the eigenvalues. We use the exact eigenvalues given by \autoref{eq:eigenvalues}, from \citet{berthon2007}, whereas in \citet{gonzalez2007}, to save computational time, the eigenvalues are computed once at the beginning of the simulation and then interpolated. However, this approximation does not impact the result.

  \subsection{Shadow} \label{sect:shadow}

  Let us now consider a two-dimensional test with source terms. Following \citet{hayes2003, gonzalez2007}, we consider a shadow test. The computational domain is a cylinder of length $L = 1\ \mathrm{cm}$ and radius $R = 0.12\ \mathrm{cm}$. It is discretized with $280 \times 80$ cells. A spheroid clump is located at the center of the box, on the symmetric axis: $(z_c, r_c) = (0.5, 0)$. The extension of the clump is $(z_0, r_0) = (0.1, 0.06)$. Initially, the medium is at equilibrium with the radiation, with $T_0 = T_r = 290\ \mathrm{K}$. We consider a homogeneous gas, with $\rho_0 = 1\ \mathrm{g\ cm^{-3}}$, except for the clump with density $\rho_1 = 100 \rho_0$. The boundary of the clump is smoothed: $\rho(r, z) = \rho_0 + \frac{\rho_1-\rho_0}{1+\exp \Delta}$ with $\Delta = 10 \left( \left( \frac{z-z_c}{z_0} \right)^2 + \left( \frac{r-r_c}{r_0} \right)^2 -1 \right)$. The opacity in the medium is $\sigma = \sigma_0 \left( \frac{T}{T_0} \right)^{-3.5} \left( \frac{\rho}{\rho_0} \right)^2$ with $\sigma_0 = 0.1\ \mathrm{cm^{-1}}$. At time $t=0$, a source is lighted at the left boundary with $T_r = 1740\ \mathrm{K}$ and the reduced flux is set to $f = 1$. Because $f$ is close to $1$ in the free-streaming regime, we encounter $f > 1$ in the simulation. To tackle this issue, we use the non-well-balanced scheme: the radiative flux source term is discretized using \autoref{eq:wb_naive}. Because we are in the free-streaming regime, we cannot use large time steps. For performance reasons, we use the semi-implicit scheme. To recover the same result as in \citet{gonzalez2007}, we use $\lambda_{i+\frac{1}{2}}^+ = \max(0.1 \times c, \lambda_{max})$ and $\lambda_{i+\frac{1}{2}}^- = \min(-0.1 \times c, \lambda_{min})$, where $\lambda_{max}$ and $\lambda_{min}$ are given by \autoref{eq:eigenvalues}.

  \begin{figure}
    \centering
    \includegraphics{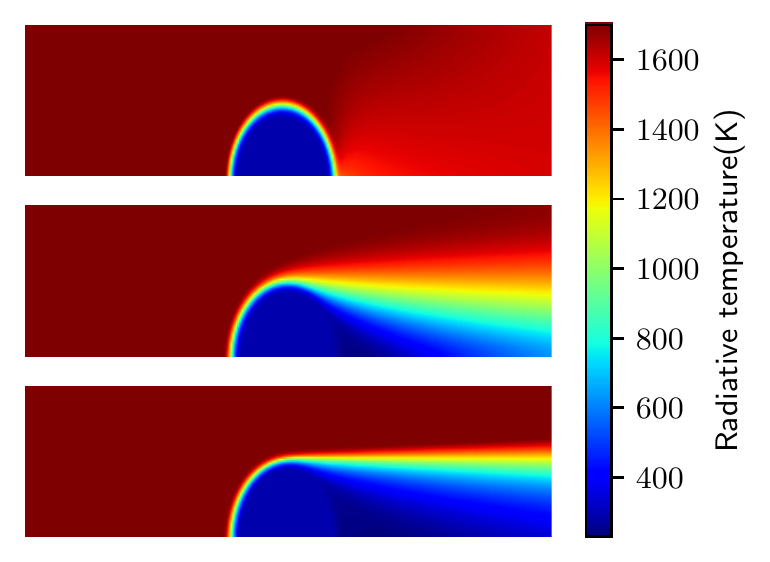}
    \caption{Shadow simulation. Snapshots of the radiative temperature at time $t_f = 10^{-10}\ \mathrm{s}$ with different closure relations: P\textsubscript{1} model (upper panel), M\textsubscript{1} model with fixed eigenvalues (middle panel) and M\textsubscript{1} model with computed eigenvalues (lower panel).}
    \label{fig:shadow}
  \end{figure}

  \Autoref{fig:shadow} shows the radiative temperature at the final time $t_f = 10^{-10}\ \mathrm{s}$ with different closure relations: the P\textsubscript{1} model, the M\textsubscript{1} model with fixed eigenvalues and the M\textsubscript{1} model with computed eigenvalues. Because of the high opacity in the clump, the light does not cross it and we expect the shadow behind it to remain stable.

  \begin{figure}
    \centering
    \includegraphics{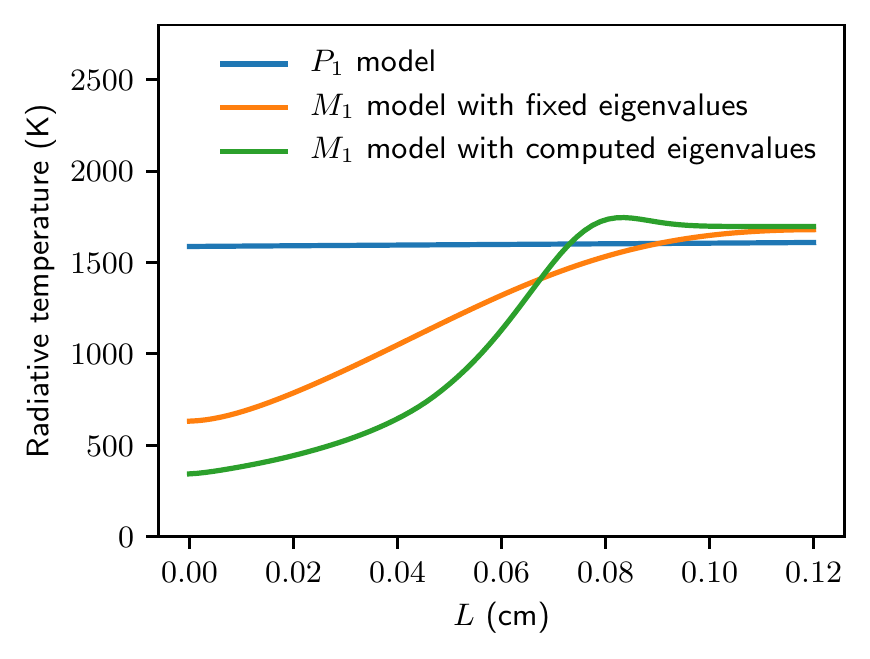}
    \caption{Shadow simulation. The figure shows the radial profiles of the radiative temperature at time $t_f = 10^{-10}\ \mathrm{s}$ with different closure relations: P\textsubscript{1} model, M\textsubscript{1} model with fixed eigenvalues and M\textsubscript{1} model with computed eigenvalues.}
    \label{fig:shadowProfile}
  \end{figure}

  As in \citet{hayes2003, gonzalez2007}, we plot the radial profile of the radiative temperature at the right boundary (\autoref{fig:shadowProfile}). Using the P\textsubscript{1} model, the radiative pressure is isotropic, therefore the photons go around the obstacle immediately, heating the whole domain. Using the M\textsubscript{1} closure relation, the shadow is better preserved, the temperature behind the obstacle remains at its initial value, $290\ \mathrm{K}$. As the opacity remains rather low outside of the clump and the light has not crossed the obstacle, the asymptotic correction has no impact on the result. Because the boundary of the clump is smoothed, the transition between the optically thick and thin medium is less sharp than in \autoref{sect:wbTest} and the well-balanced modification of the source term is not necessary.

  \subsection{Radiative shocks} \label{test:shocks}

  We finally consider radiative shocks: the gas and the radiation exchange energy and momentum. Following \citet{ensman1994, hayes2003, gonzalez2007}, we consider a one-dimensional homogeneous medium, with $\rho = 7.78 \times 10^{-10}\ \mathrm{g\ cm^{-3}}$ and $\sigma = 3.1 \times 10^{-10}\ \mathrm{cm^{-1}}$. We consider a perfect gas with an adiabatic coefficient $\gamma = \frac{7}{5}$ and a mean molecular weight $\mu = 1$. The length of the domain is $7 \times 10^{10}\ \mathrm{cm}$. It is discretized with $400$ cells. The initial temperature at the left boundary is set to $10\ \mathrm{K}$ and it is increased by $0.25\ \mathrm{K}$ per cell. Initially, the radiation is at equilibrium with the gas. The left boundary condition is reflective, the initial velocity of the fluid is set to $u_0$. According to the value of $u_0$, the shock will be subcritical or supercritical. See \citet{gonzalez2007} for more details. To compare our results with \citet{ensman1994, hayes2003, gonzalez2007}, we plot the temperature as a function of $x_i = x - u_0t$.

  \subsubsection{Subcritical shock} \label{sect:subcriticalShock}

  \begin{figure}
    \centering
    \includegraphics{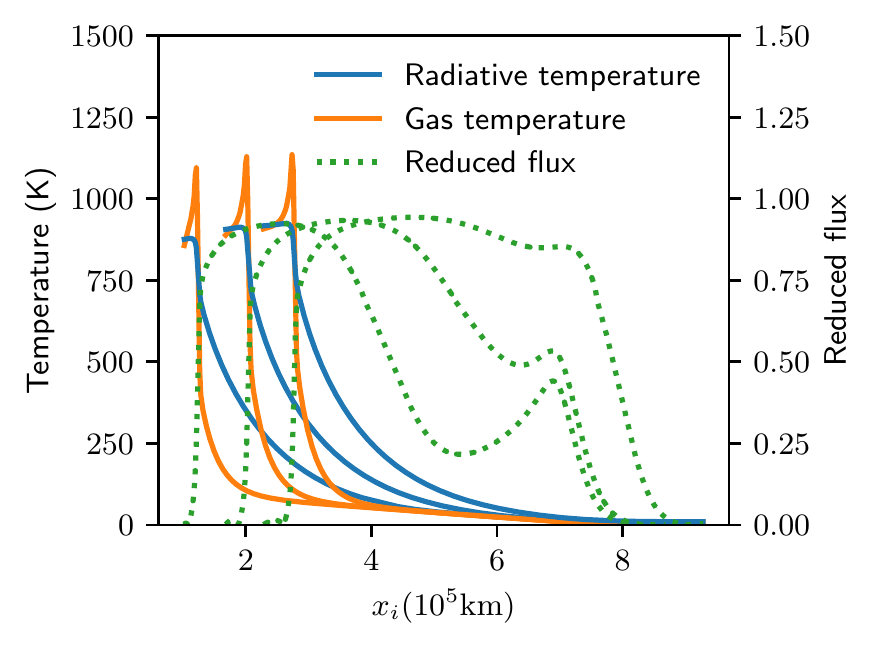}
    \caption{Subcritical shock simulation. The figure shows snapshots of gas temperature, radiative temperature, and reduced flux at different times: $1.7 \times 10^4\ \mathrm{s}$, $2.8 \times 10^4\ \mathrm{s}$, and $3.8 \times 10^4\ \mathrm{s}$.}
    \label{fig:subcriticalShock}
  \end{figure}

  We first consider a subcritical shock, the initial velocity is set to $u_0 = -6\ \mathrm{km\ s^{-1}}$. \Autoref{fig:subcriticalShock} shows the gas temperature, the radiative temperature, and the reduced flux at three different times: $1.7 \times 10^4\ \mathrm{s}$, $2.8 \times 10^4\ \mathrm{s}$, and $3.8 \times 10^4\ \mathrm{s}$. As expected, the gas and the radiation are not at equilibrium, before nor after the shock. The gas temperature reaches $1135\ \mathrm{K}$, as in \citet{gonzalez2007}, whereas it is only $850\ \mathrm{K}$ in \citet{ensman1994}.

  \subsubsection{Supercritical shock} \label{sect:supercriticalShock}

  \begin{figure}
    \centering
    \includegraphics{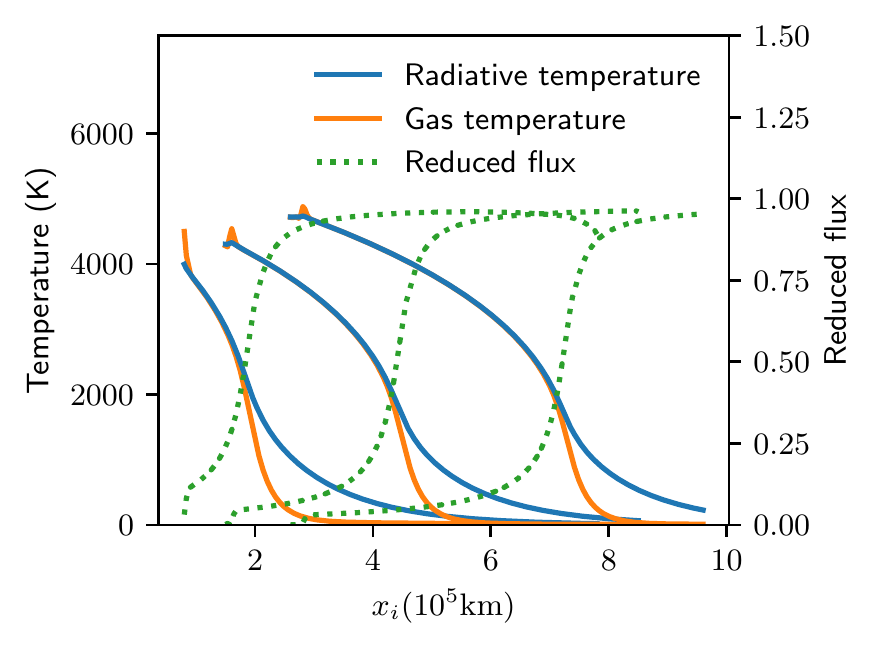}
    \caption{Supercritical shock simulation. The figure shows snapshots of gas temperature, radiative temperature, and reduced flux at different times: $4.0 \times 10^3\ \mathrm{s}$, $7.5 \times 10^3\ \mathrm{s}$, and $1.3 \times 10^4\ \mathrm{s}$.}
    \label{fig:supercriticalShock}
  \end{figure}

  We consider now a supercritical shock, where the initial velocity is set to $u_0 = - 20\ \mathrm{km\ s^{-1}}$. \Autoref{fig:supercriticalShock} shows the gas temperature, the radiative temperature, and the reduced flux at three different times: $4 \times 10^3\ \mathrm{s}$, $7.5 \times 10^3\ \mathrm{s}$, and $1.3 \times 10^4\ \mathrm{s}$. As in \citet{gonzalez2007}, the radiative temperature is the same as the matter temperature on both sides of the shock. The gas and the radiation are therefore at equilibrium. The radiative precursor is larger than the subcritical shock's radiative precursor, as intended, and the temperature reaches $5\,000\ \mathrm{K}$, as in \citet{ensman1994}. We also recover the Zel'dovich spike.

  \subsection{Expansion of \ion{H}{ii} region}\label{sect:hiiRegionExpansion}

  Now that we confirmed the good behavior of the numerical scheme with both the asymptotic preserving and the well-balanced properties, we can apply it to a physical situation: the propagation of the ionization front in a massive pre-stellar dense core.

  \subsubsection{Model} \label{sect:hiiModel}

  We consider the early stage of the development of an \ion{H}{ii} region in a massive pre-stellar dense core \citep{churchwell2002}. We focus on a region of the dense core at about $100\ \mathrm{AU}$ from the massive young stellar object (YSO). This region has been heated by the YSO during the pre-main sequence phase. The temperature reached at this location by infrared heating is of the order of $1\, 000\ \mathrm{K}$ and the transport of energy in this region can be dominated by convection. We have inferred the convective state of this region by computing thermal and adiabatic gradients based on observations of \citealt{herpin2009} (Fig. 7). High-energy photons emitted by the YSO when entering the main sequence will start to ionize the surrounding gas. This will trigger the propagation of an ionization front in a convective medium, and we are interested in the stability of such a front perturbed by the pre-existing convective motions. 
  
  The interaction of the ionizing photons with the gas is described by \autoref{eq:radiativeHydrodynamicsEqua}. The only photons able to ionize the gas are emitted by the YSO, i.e. there is no local source of ionizing photons. Following \citet{tremblin2012}, we need to modify this model to take into account photo-chemistry and thermal balance. We define the fraction of ionization $X = n_{H^+} / n_H$ where $n_{H} = n_{H^+} + n_{H^0}$, $n_{H^+}$ is the number of ionized atoms and $n_{H^0}$ is the number of cold atoms. The evolution of the number of ionized atoms is just the number of incoming photons that interact with the gas minus the number of ionized atoms that recombine (on the spot approximation, see \citealt{lesaffre2002}). Therefore, 
  \begin{equation}
    \partial_t (\rho X) + \nabla \cdot (\rho X \vec{U}) = \sigma_{\gamma} F_{\gamma} n_H (1-X) - \beta X^2 n_H^2,
  \end{equation}
  where $F_{\gamma}$ is the number of incoming photons per unit of surface and time, $\sigma_{\gamma}$ is the average cross-section at the temperature of the star and  $\beta$ gives the recombination rate: $\beta = 2 \times 10^{-10} T^{-0.75}$ with $T$ the temperature of thermodynamic equilibrium \citep{black1981}.

  The thermal balance is the difference between the heating rate and the cooling rate. The extra energy of the absorbed photons is converted into kinetic energy of electrons. It is the only source of heating during the ionization, hence the heating rate is given by $(1-X)n_H F_{\gamma} \sigma_{\gamma} e_{\gamma}$. In this simplified model, the equilibrium temperature is obtained from the balance between the heating from the ionization and the cooling from the recombination. We do not consider any other effects such as metal cooling. Therefore, we take $e_{\gamma} = 1\ \mathrm{eV}$ \citep{lesaffre2002} to recover the observed temperature around $1\, 000\ \mathrm{K}$. From \citet{tremblin2012}, the cooling rate is given by $\beta X^2 n_{H}^2 k_b T / (\gamma-1)$. We also add a term of Newtonian forcing: $\partial_t T_g = \frac{T_g - T_{forcing}}{\tau_{forcing}}$ to trigger convection. $T_{forcing}$ is the equilibrium temperature profile, depending on space, and $\tau_{forcing}$ is the relaxation timescale. The gas temperature will relax toward the equilibrium temperature profile $T_{forcing}$.

  By writing $c E_r = F_{\gamma} e_{\gamma}$, $\rho = n_H m_H$ and $\boldsymbol\sigma = \boldsymbol\sigma_{\boldsymbol\gamma} n_H$, we finally have to solve the following system:
  \begin{equation}
    \left\{
      \begin{aligned}
        \partial_t \rho + \nabla \cdot (\rho \vec{u}) &= 0\\
        \partial_t (\rho \vec{u}) + \nabla \cdot (\rho \vec{u} \otimes \vec{u} + p \tens{I}) &= \rho \vec{g} + \frac{\sigma(1-X)}{c} \vec{F}_r\\
        \partial_t (\rho E) + \nabla \cdot ((\rho E+p) \vec{u}) &= \rho \vec{g} \cdot \vec{u} +c \sigma(1-X) E_r \\
        &- \beta \frac{\rho^2 X^2}{m_H^2} \frac{k_b T_g}{\gamma-1} - \frac{T_g - T_{forcing}}{\tau_{forcing}} \\
        \partial_t E_r + \nabla \cdot \vec{F}_r &= - c \sigma (1-X) E_r \\
        \partial_t \vec{F}_r + \nabla \cdot \tens{P}_r &= - c \sigma(1-X) \vec{F}_r\\
        \partial_t (\rho X) + \nabla \cdot (\rho X \vec{u}) &= \frac{\sigma (1-X) c E_r m_H}{e_{\gamma}} - \frac{\beta \rho^2 X^2}{m_H}.
      \end{aligned}
    \right.
  \end{equation}

  In this test, we will use the M\textsubscript{1} solver with the asymptotic correction presented in \autoref{sect:ap}, but we do not use the well-balanced discretization of the source term because of stability issues that will be discussed in \autoref{sect:concwb}.

  \subsubsection{Setup} \label{sect:hiiSetup}

  We consider a square domain with a side $5\ \mathrm{AU}$ long. We use a setup similar to \citet{padioleau2019} for compressible convection simulations. The temperature is set at the top and the bottom of the box at $500\ \mathrm{K}$ and $1\,000\ \mathrm{K}$ respectively. The initial temperature is a linear interpolation between the top and the bottom of the box. It is also the forcing temperature profile $T_{forcing}$. We take $\tau_{forcing} = 10^7\ \mathrm{s}$. These parameters are chosen to trigger the initial convective motions. We also set the pressure at the bottom: $10^{-3}\ \mathrm{dyne \cdot cm^{-2}}$ \citep{herpin2009}. The density and the pressure are linked by the ideal gas law: $p = \frac{\rho k_b T_g}{m_H \mu}$, where $\mu$ is the mean molecular weight. The non-ionized medium is made of hydrogen, with $\mu_1=1$. When the medium is fully ionized, it is made of atomic nucleus and electrons, so twice as many particles for the same mass. Because the distribution of nucleus and electrons is homogeneous, the mean molecular weight is $\mu_2=0.5$. When the medium is partially ionized, we take $\mu$ as the mean of the previous values balanced by the fraction of ionization, i.e. $\mu = (1-X) \mu_1 + X \mu_2$. The density is initialized with the recursive formula $p_{i+1} - p_{i} = \frac{1}{2} \left( \rho_i + \rho_{i+1} \right) g \Delta z$, which is the discrete version of the hydrostatic balance $\nabla p = - \rho \vec{g}$.
  
  We impose Neumann boundary conditions for the temperature. The pressure and density are imposed by an extrapolation of the hydrostatic balance. Because the hydrodynamics solver is well-balanced for the gravity, this configuration will remain static, even if the initial condition is unstable. The hydrostatic equilibrium is destabilized with a velocity mode perturbation of the form
  \begin{equation}
    \begin{aligned}
      u(x, y) &= 2 \cdot 10^{-4} c_s \sin\left(2 \pi \frac{x-x_{mid}}{L_x}\right) \sin\left( \pi \frac{y-y_{mid}}{L_y} \right)\\
      v(x, y) &= 2 \cdot 10^{-4} c_s \cos\left(2 \pi \frac{x-x_{mid}}{L_x}\right) \cos\left( \pi \frac{y-y_{mid}}{L_y} \right),
    \end{aligned}
  \end{equation}
  with $c_s$ the speed of sound, $x_{mid} = y_{mid} = 2.5\ \mathrm{AU}$ and $L_x = L_y = 5\ \mathrm{AU}$. Without any interaction with the ionizing photons, the convective motions are stationary.

  The opacity is set to $\sigma = \frac{\sigma_{\gamma} \rho}{m_H}$ with $\sigma_{\gamma} = 6 \times 10^{-18}\ \mathrm{cm^2}$ \citep{lesaffre2002}. The radiative energy and flux are set to $0$ and the medium is not ionized ($X = 0$). We initialize the  hydrodynamics variables with the steady state described previously. At time $t = 0$, the bottom boundary of the region is ionized: the reduced flux is set to $1$ and the radiative energy is set to $\frac{F_* e_{\gamma}}{c}$ with $F_* = 3 \times 10^{17}\ \mathrm{cm^{-2}\ s^{-1}}$ in the boundary. The boundary conditions for the hydrodynamics variables remain unchanged.

  \subsubsection{Results} \label{sect:hiiResults}

  \begin{figure}
    \centering
    \includegraphics{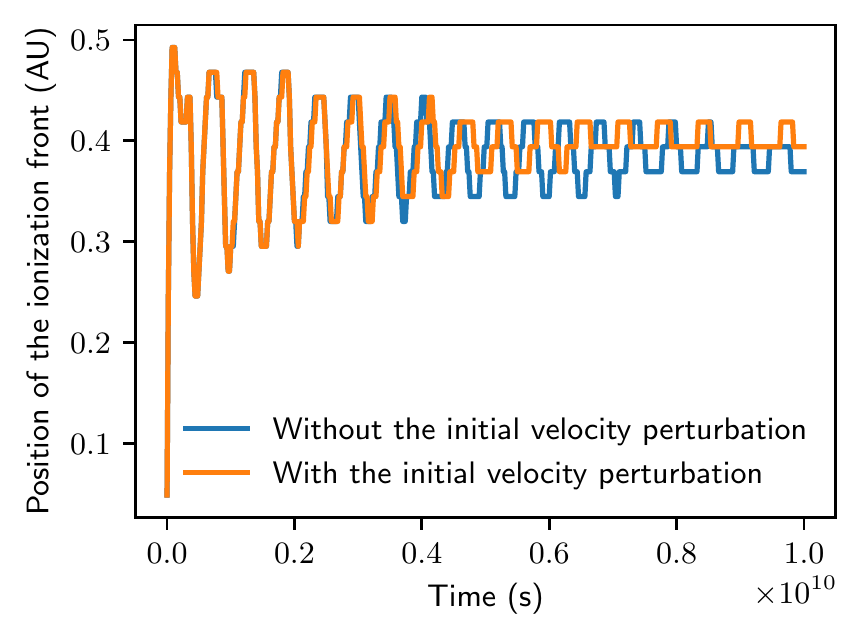}
    \caption{Evolution of the position of the ionization front as a function of time, with and without the initial velocity perturbation.}
    \label{fig:hii_evolution}
  \end{figure}

  As the initial condition is such that $E_r$ is close to $0$, this can easily create some spurious values such that $f > 1$. This is clearly a numerical artifact induced by the very low value of the radiative energy in regions where no ionizing photons are present. Even with a centered discretization of the radiative flux source term and without the asymptotic correction, we still encounter $f > 1$ during the simulation. Because of this problem, we impose $f=1$ in the  computation of the Eddington tensor in the cells where $f > 1$.

  \Autoref{fig:hii_evolution} shows the evolution of the position of the ionization front as a function of time. With and without the initial convective rolls, the position of the ionization front oscillates around an equilibrium position, between $0.3\ \mathrm{AU}$ and $0.4\ \mathrm{AU}$. The oscillations around the equilibrium are expected and have been observed with simpler models \citep{tremblin2012-article}.

  \begin{figure*}
    \centering
    \includegraphics{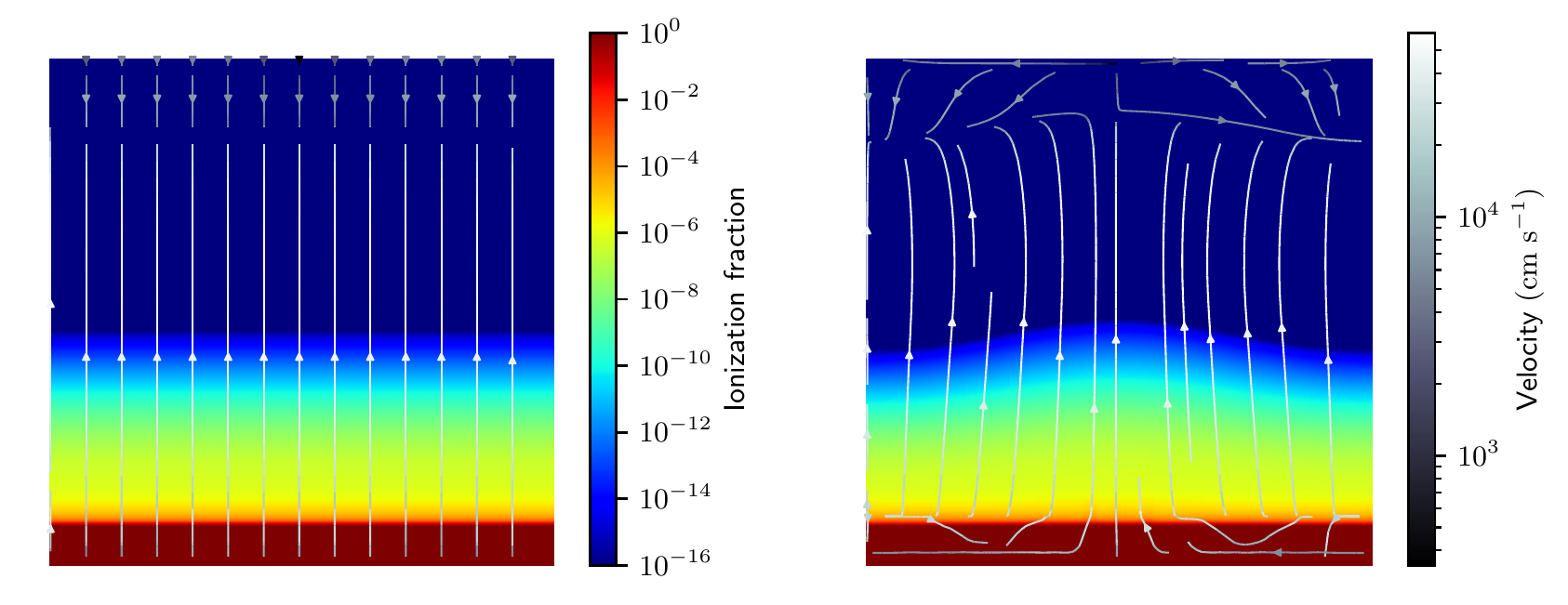}
    \caption{Snapshots of the fraction of ionization and the velocity field at time $t = 6 \times 10^{8}\ \mathrm{s}$ without the initial velocity perturbation (left panel) and with it (right panel).}
    \label{fig:hii_region_short}
  \end{figure*}

  \begin{figure*}
    \centering
    \includegraphics{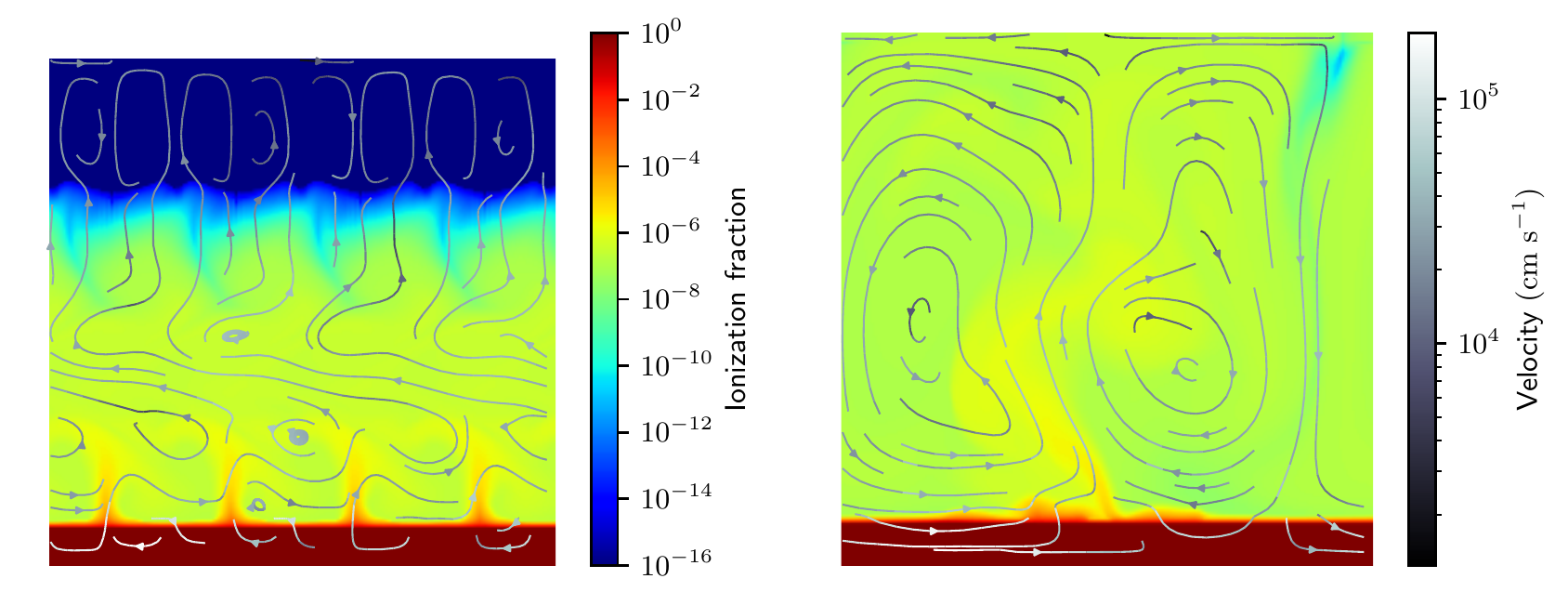}
    \caption{Snapshots of the fraction of ionization and the velocity field at the final time $t_f = 10^{10}\ \mathrm{s}$ without the initial velocity perturbation (left panel) and with it (right panel).}
    \label{fig:hii_region}
  \end{figure*}

  \Autoref{fig:hii_region_short} and \autoref{fig:hii_region} show the ionization front at time $t = 6 \times 10^{8}\ \mathrm{s}$ and at the final time $t_f = 10^{10}\ \mathrm{s}$. With and without the initial convective rolls, a numerical noise appears, as a consequence of the long timescales. Because of the numerical noise, some lack of symmetry can appear, such as on the left panel of \autoref{fig:hii_region_short}. The fraction of ionization, which is always between $0$ and $1$, reaches values between $10^{-12}$ and $10^{-6}$. The effect of the preconditioner and the MPI domain decomposition is discussed in \autoref{appendix:hii_region}. However, the numerical noise does not affect the position of the ionization front.

  The stability of the ionization front is an issue that has been discussed for a long time in the literature \citep{mizuta2005}. For example, 3D simulations of the expansion of a spherical ionization front in 3D Cartesian grids have shown instabilities either on the axis of the grid or in the diagonal depending on the numerical scheme (see Fig. A3 \citealt{bisbas2015}). The dependence of the instability on the grid cast doubts about a possible physical regime. Our test case shows that even with convective motions of large amplitude, the ionization front remains stable.

  \section{Discussion and conclusion} \label{sect:conclusion}

  We have presented a new radiation hydrodynamics code. The radiative transfer is described with a two-moment model and uses the M\textsubscript{1} closure relation. We first discuss some limitations of our work before reaching our conclusion.

  \subsection{Well-balanced discretization of the source term} \label{sect:concwb}

  In \autoref{sect:wb} we have proposed a well-balanced discretization of the source term on the  radiative flux equation. This discretization allows us to properly capture the steady state with constant flux and with a discontinuity of opacity. However, this discretization can lead to spurious oscillations in the radiative flux, a problem that we have encountered in the test case for the expansion of \ion{H}{ii} regions. Although we have changed the discretization of the source term to achieve a well-balanced property, our integration of the hyperbolic part on the source term is still split into two steps. Such a splitting strategy might be unstable if the source term is not taken into account in the hyperbolic part. A possible solution to this problem would be to incorporate the source term in a Lagrange-projection-like scheme such as \citet{buet2008}. Such a strategy might be necessary to treat the radiation hydrodynamics problem of cloud interfaces.

  \subsection{Asymptotic limit for radiation hydrodynamics} \label{sect:apVelocity}

  In \autoref{sect:ap}, we have presented an asymptotic correction which allows us to capture the asymptotic behavior, whereas the solution given by a standard scheme is dominated by numerical diffusion. The asymptotic correction uses the numerical diffusion to recover the physical one, in a static fluid. Nevertheless, this scheme does not capture the asymptotic regime in a moving fluid, as presented in \autoref{appendix:diffusiveLimitRHD}. Most of the schemes proposed in the literature do not preserve this asymptotic regime (e.g. \citealt{gonzalez2007, berthon2011}). The diffusive regime depends on the material velocity, our scheme cannot reach it. A possible solution would be to limit the numerical diffusion with a correction similar to a low Mach correction, e.g. \citet{chalons2016}, in conjunction with a cell-centered discretization of the source term, as proposed by the Lagrange-projection scheme of \citet{buet2008}.

  \subsection{Conclusion}

  The model for radiation hydrodynamics proposed in this paper has the correct behavior in both free-streaming and diffusive limits. It is discretized with an asymptotic preserving scheme. This asymptotic correction is important to capture the correct behavior in the diffusive limit and to preserve the free-streaming regime in a static fluid. 
  
  We take advantage of the libraries Kokkos and Trilinos to reach high-performance computing and to solve linear systems. This approach allows us to take advantage of different architectures and to use large time steps for radiative transfer. Using the implicit solver is profitable as soon as the time step given by the hydrodynamics is $50-100$ times larger than the explicit time step for radiative transfer, depending on the preconditioner and the architecture. 
  
  The solver is then coupled with the hydrodynamics code that implements an all-regime and well-balanced solver for hydrodynamics with gravity. The tests performed show that ARK-RT is well suited to study many astrophysical problems. Using this code, we have been able to study the development of \ion{H}{ii} regions in massive pre-stellar dense cores, especially the propagation of the ionization front in the presence of convection. We have shown that even with the destabilizing effect of convection, the ionization front is strongly stable against perturbations. A linear stability analysis similar to the Rayleigh-Taylor instability but including source term could give more insight into this behavior.

  Further work will consist of the development of a numerical scheme that preserves the admissible states while keeping the asymptotic preserving and well-balanced properties. We will then be able to take full advantage of the largest next generation massively parallel architectures to study, among others, atmospheric physics with opacity interfaces.

  \begin{acknowledgements}
    PT acknowledges supports by the European Research Council under Grant Agreement ATMO 757858. The authors thank the referee for his useful comments, especially about the test on the expansion of \ion{H}{ii} regions. HB thanks also Rodolphe Turpault for his welcome in Bordeaux and the helpful discussion.
  \end{acknowledgements}

  \bibliographystyle{aa} 
  \bibliography{main}

  \begin{appendix}
  \section{Diffusive limit for radiation hydrodynamics} \label{appendix:diffusiveLimitRHD}

  As in \autoref{sect:diffusiveLimit}, we consider the diffusive limit with the P\textsubscript{1} closure relation. We introduce a rescaling parameter $\varepsilon$ to write the time (resp. the opacity) as $\tilde{t} = \varepsilon t$ (resp. $\tilde{\sigma} = \varepsilon \sigma$). Because the velocity of the fluid is smaller than the speed of light ($\frac{u}{c} << 1$), we also rescale it as $\vec{\tilde{u}} = \frac{\vec{u}}{\varepsilon}$. Let us focus on the equations describing the evolution of the radiative variables:
  \begin{subequations}
    \begin{empheq}[left=\empheqlbrace]{align}
      \notag \varepsilon^2 \partial_{\tilde{t}} E_r + \varepsilon \nabla \cdot \vec{F}_r &  = c \tilde{\sigma} \left( a_r T_g^4 - E_r \right) \\\begin{aligned}[t]
      \end{aligned}
      &+ \varepsilon \frac{\tilde{\sigma}}{c} \vec{\tilde{u}} \cdot \vec{F}_r + \varepsilon^2 \frac{4}{3} \frac{\tilde{\sigma}}{c} E_r \vec{\tilde{u}} \cdot \vec{\tilde{u}} \label{eq:asymptoticDevRHD:energy}\\
      \varepsilon^2 \partial_{\tilde{t}} \vec{F}_r + \varepsilon \frac{c^2}{3} \nabla E_r &=  -\tilde{\sigma} c \vec{F}_r + \varepsilon \frac{c}{3} \tilde{\sigma} \vec{\tilde{u}} E_r\notag\\
      & + \varepsilon c \tilde{\sigma} \vec{\tilde{u}} a_r T_g^4 + \varepsilon^2 \frac{\tilde{\sigma}}{c} \vec{\tilde{u}} \cdot (\vec{\tilde{u}} \cdot \vec{F}_r). \label{eq:asymptoticDevRHD:flux}
    \end{empheq}
  \end{subequations}
  By expanding \autoref{eq:asymptoticDevRHD:energy} and \autoref{eq:asymptoticDevRHD:flux} at order $0$, we have
  \begin{equation}
    \left\{
      \begin{aligned}
        E_{r,0} &=  a_r T_{g,0}^4\\
        \vec{F}_{r,0} &= \vec{0}.
      \end{aligned}
    \right.
  \end{equation}
  Expanding \autoref{eq:asymptoticDevRHD:flux} at order $1$ leads to
  \begin{equation}
    \vec{F}_{r,1} = - \frac{c}{3 \tilde{\sigma}} \nabla E_{r,0} + \frac{4}{3} E_{r,0} \vec{\tilde{u}}_{0}.
  \end{equation}

  Finally, looking at the radiative energy and the gas internal energy at order $2$, source terms cancel each other, only the divergence of the radiative flux at order $1$ remains, and we have
  \begin{equation}
    \partial_{\tilde{t}} \left( \rho c_v T_{g,0} + E_{r,0} \right) - \nabla \cdot \left( \frac{c}{3 \tilde{\sigma}} \nabla E_{r,0} \right) = - \frac{4}{3} \nabla \left( E_{r,0} \vec{\tilde{u}}_0 \right).
  \end{equation}
  One can also look at \autoref{eq:asymptoticDevRHD:energy} at order $2$. Expanding \autoref{eq:asymptoticDevRHD:energy} at order $2$ gives
    \begin{equation}
      \begin{aligned}
        \partial_{\tilde{t}} E_{r,0} - \nabla \left( \frac{c}{3 \tilde{\sigma}} \nabla E_{r,0} \right) &= c \tilde{\sigma} \left( 6 a_r T_{g,0}^2 T_{g,1}^2 + 4 a_r T_{g,0}^3 T_{g,2} - E_{r,2} \right)\\
        & - \frac{4}{3} \nabla \cdot \left( E_{r,0} \vec{u}_0 \right) - \frac{1}{3} \vec{u}_0 \cdot \nabla E_{r,0} + 2 \frac{4}{3} \frac{\sigma}{c} E_{r,0} \vec{u}_0^2.\label{eq:diffusionEquaRHD}
      \end{aligned}
    \end{equation}
    We recover Eq. 43 of \citet{krumholz2007}. The term $c \tilde{\sigma} \left( 6 a_r T_{g,0}^2 T_{g,1}^2 + 4 a_r T_{g,0}^3 T_{g,2} - E_{r,2} \right)$ is the development at second order of the term $\kappa_0 (4\pi B - c E)$. Because we do not neglect any terms $O\left( \frac{\vec{\tilde{u}}}{c} \right)$, some coefficients are slightly different. See the discussion in \citet{krumholz2007} for the importance of the term $\frac{4}{3} \frac{\tilde{\sigma}}{c} E_{r,0} \vec{\tilde{u}}_0^2$.

    \section{Von Neumann stability analysis for the well-balanced modification of the source term} \label{appendix:wbStability}

    For simplicity, we split \autoref{eq:momentModel:flux} into a pure hyperbolic problem $\partial_t \vec{F}_r + c^2 \nabla \cdot \tens{P}_r = 0$ and a source problem $\partial_t \vec{F}_r = -c \sigma \vec{F}_r$. We focus on the one-dimensional source problem, with periodic boundary conditions on the domain $[0, T] \times [0, 1]$ with $T$ the final time. The following can easily be extended to an arbitrary space interval. Because we use periodic boundary conditions, we can apply the von Neumann stability analysis (see e.g. \citealt{anderson1995}), based on the decomposition of the numerical solution into Fourier series. Let us recall that, using the well-balanced modification of the source term, the source problem is discretized as
      \begin{equation}
        F_j^{n+1} + r_{j-\frac{1}{2}} F_{j-1}^{n+1} + \left( r_{j-\frac{1}{2}} + r_{j+\frac{1}{2}} \right) F_j^{n+1} + r_{j+\frac{1}{2}} F_{j+1}^{n+1} = F_j^{n},
        \label{eq:schemeStabilityAnalysis}
      \end{equation}
      with $r_{j+\frac{1}{2}} = \frac{c \sigma_{j+\frac{1}{2}} \Delta t}{4}$. We define the function $F^n$, piecewise constant, such that
      \begin{equation}
         F^n(x) = 
          \left\{
            \begin{aligned}
              F_j^n \text{ if } x_{j-\frac{1}{2}} < x < x_{j+\frac{1}{2}}\\
              0 \text{ otherwise.}
            \end{aligned}
          \right.
      \end{equation}
      This function is then extended to $\mathbb{R}$ by periodicity.  $F^n$ can now be expanded in a Fourier series:
      \begin{equation}
        F^n(x) = \sum_{k \in \mathbb{Z}} \hat{F}^n(k) e^{2 i k \pi x},
      \end{equation}
      with 
      \begin{equation}
        \hat{F}^n(k) = \int_0^1 F^n(x) e^{- 2 i k \pi x} dx.
      \end{equation}
      We can define the $2$-norm of the function $F^n$:
      \begin{equation}
        \begin{aligned}
          ||F^n||_2 &= \left( \int_0^1 \left( F^n(x) \right)^2 dx \right)^{\frac{1}{2}}\\
          &= \left( \sum_{k \in \mathbb{Z}} |\hat{F}^n(k)|^2 \right)^{\frac{1}{2}}\\
          &= \left( \sum_{j=1}^J \Delta x\left( F_j^n \right)^2 \right)^{\frac{1}{2}}.
        \end{aligned}
      \end{equation}
      We apply the Fourier transform to \autoref{eq:schemeStabilityAnalysis}:
      \begin{equation}
        \hat{F}^{n+1}(k) \left( 1 + r_{j-\frac{1}{2}} e^{- 2 i k \pi \Delta x} + r_{j-\frac{1}{2}} + r_{j+\frac{1}{2}} + r_{j+\frac{1}{2}} e^{2 i k \pi \Delta x} \right) = \hat{F}^n(k).
      \end{equation}
      We define the amplification factor $A(k)$ as
      \begin{equation}
        A(k) = \frac{1}{1 + r_{j-\frac{1}{2}} e^{- 2 i k \pi \Delta x} + r_{j-\frac{1}{2}} + r_{j+\frac{1}{2}} + r_{j+\frac{1}{2}} e^{2 i k \pi \Delta x}},
      \end{equation}
      and we have then $\hat{F}^{n+1}(k) = A(k) \hat{F}^n(k)$. By induction, we have $\hat{F}^{n}(k) = \left(A(k)\right)^n \hat{F}^0(k)$. The coefficient $\hat{F}^n(k)$ remains bounded if and only if $|A(k)| \le 1$. In this case, for all $k \in \mathbb{Z}$, $|\hat{F}^{n+1}(k)| \le |\hat{F}^n(k)|$. Therefore, $||F^{n+1}||_2 \le ||F^n||_2 \le ||F^0||_2$ and the scheme is unconditionally stable. We now have to prove that $|A(k)| \le 1$:
      \begin{equation}
        \begin{aligned}
          \frac{1}{|A(k)|^2} &= \left( 1 +  r_{j-\frac{1}{2}} + r_{j+\frac{1}{2}} \right)^2 + r_{j-\frac{1}{2}}^2 + r_{j+\frac{1}{2}}^2 \\
          &+ 2 r_{j-\frac{1}{2}} \left( 1 + r_{j-\frac{1}{2}} + r_{j+\frac{1}{2}} \right)  \cos(2 k \pi \Delta x)\\
          &+ 2 r_{j+\frac{1}{2}} \left( 1 + r_{j-\frac{1}{2}} + r_{j+\frac{1}{2}} \right) \cos (2 k \pi \Delta x)\\
          &+ 2 r_{j-\frac{1}{2}} r_{j+\frac{1}{2}} \cos(4 k \pi \Delta x) \\
          & = \left( 1 + \underbrace{r_{j-\frac{1}{2}} \left(1 - \cos(2 k \pi \Delta x)\right)}_{\ge 0} + \underbrace{r_{j+\frac{1}{2}} \left(1 - \cos(2 k \pi \Delta x)\right)}_{\ge 0} \right)^2\\
          &+ \underbrace{\left( r_{j-\frac{1}{2}} \sin(2 k \pi \Delta x) - r_{j+\frac{1}{2}} \sin(2 k \pi \Delta x)\right)^2}_{\ge 0}\\
          & \ge 1.
        \end{aligned}
      \end{equation}
      Since $\frac{1}{|A(k)|^2} \ge 1$, we have $|A(k)| \le 1$.
    \section{Numerical scheme in the diffusive limit} \label{appendix:proofAP}
    We consider the numerical scheme developed in \autoref{sect:numericalScheme} in the asymptotic regime, with $\sigma_{i+\frac{1}{2}} \Delta x \rightarrow \infty$. Following \autoref{sect:diffusiveLimit}, we introduce the rescaling parameter $\varepsilon$ to write the time (resp. the opacity) as $\tilde{\Delta t} = \varepsilon \Delta t$ (resp. $\tilde{\sigma} = \varepsilon \sigma$). Using the P\textsubscript{1} closure relation, we have $\lambda_{i+\frac{1}{2}}^+ = - \lambda_{i+\frac{1}{2}}^- = \frac{c}{\sqrt{3}}$ and
    \begin{subequations}
      \begin{align}
        \notag \varepsilon^2 E_{i}^{n+1} &= \varepsilon^2 E_{i}^n - \frac{\tilde{\Delta t}}{\Delta x} \left( \varepsilon \alpha_{i+\frac{1}{2}} \mathcal{F}^*_{i+\frac{1}{2}} - \varepsilon \alpha_{i-\frac{1}{2}} \mathcal{F}^*_{i-\frac{1}{2}} \right) \\
        &+ c \tilde{\Delta t} \tilde{\sigma}_i \left( a_r \left( T_{i}^{n+1} \right)^4 - E_{i}^{n+1} \right) \label{eq:numericalSchemeAsymptoticDev:energy}\\
        \notag \varepsilon^2 F_{i}^{n+1} &= \varepsilon^2 F_{i}^n - \frac{\tilde{\Delta t}}{\Delta x} \left(\varepsilon \mathcal{P}^*_{i+\frac{1}{2}} - \varepsilon \mathcal{P}^*_{i-\frac{1}{2}} \right) \\
        &- \frac{c \tilde{\Delta t}}{2} \left( \tilde{\sigma}_{i+\frac{1}{2}} F_{i+\frac{1}{2}}^{n+1} + \tilde{\sigma}_{i-\frac{1}{2}} F_{i-\frac{1}{2}}^{n+1} \right) \label{eq:numericalSchemeAsymptoticDev:flux}\\
        \varepsilon^2 \rho c_v T_{i}^{n+1} &= \varepsilon^2 \rho c_v T_{i}^n - c \tilde{\Delta t} \tilde{\sigma}_i \left( a_r \left( T_{i}^{n+1} \right)^4 - E_{i}^{n+1} \right). \label{eq:numericalSchemeAsymptoticDev:temperature}
      \end{align}
    \end{subequations}
    Radiative variables are expanded, e.g. $E_i^n = E_{i,0}^n + \varepsilon E_{i,1}^n + \mathcal{O}(\varepsilon^2)$. Expanding \autoref{eq:numericalSchemeAsymptoticDev:energy} and \autoref{eq:numericalSchemeAsymptoticDev:flux} at order $0$ leads to
    \begin{equation}
      \begin{aligned}
        E_{i,0}^{n+1} &=  a_r \left( T_{i,0}^{n+1} \right)^4\\
        F_{i,0}^{n+1} &= 0.
      \end{aligned}
      \label{eq:asymptoticDevDiscreteOrder0}
    \end{equation}
    At first order for \autoref{eq:numericalSchemeAsymptoticDev:flux}, we have
    \begin{equation}
      \begin{aligned}
        \tilde{\sigma}_{i+\frac{1}{2}} F_{i+\frac{1}{2},1}^{n+1} + \tilde{\sigma}_{i-\frac{1}{2}} F_{i-\frac{1}{2},1}^{n+1} &= - \frac{c}{3} \frac{E_{i+1,0}^{n+1} - E_{i-1,0}^{n+1}}{\Delta x} \\
        &+ \underbrace{\frac{F_{i+1,0}^{n+1} - 2 F_{i,0}^{n+1} + F_{i-1,0}^{n+1}}{\sqrt{3}\Delta x}}_{ = 0}.\\
      \end{aligned}
    \end{equation}
    Using the boundary condition given by \autoref{eq:boundaryCondition}, we have
    \begin{equation}
      \tilde{\sigma}_{i+\frac{1}{2}} F_{i+\frac{1}{2}, 1}^{n+1} = - \frac{c}{3 \Delta x} \left( E_{i+1,0}^{n+1} - E_{i,0}^{n+1} \right)
      \label{eq:asymptoticDevDiscreteOrder1}
    \end{equation}
    in the whole domain.

    Now, we consider the sum of \autoref{eq:numericalSchemeAsymptoticDev:energy} and \autoref{eq:numericalSchemeAsymptoticDev:temperature} expanded at second order. If $\alpha_{i+\frac{1}{2}} = 1$, we have
      \begin{equation}
        \begin{alignedat}{2}
          \varepsilon^2 \left( E_{i,0}^{n+1} + \rho c_v T_{i,0}^{n+1} \right) = \varepsilon^2 \left( E_{i,0}^{n} + \rho c_v T_{i,0}^{n} \right)\\
          + \varepsilon \frac{\tilde{\Delta t}}{\Delta x} \frac{c}{2\sqrt{3}} \left( \alpha_{i+\frac{1}{2}} \left( E_{i+1,0}^{n+1} - E_{i,0}^{n+1} \right) - \alpha_{i-\frac{1}{2}} \left( E_{i,0}^{n+1} - E_{i-1,0}^{n+1} \right) \right)\\
          - \varepsilon^2 \frac{\tilde{\Delta t}}{2 \Delta x} \alpha_{i+\frac{1}{2}} \left( F_{i+1,1}^{n+1} + F_{i,1}^{n+1} - \frac{c}{\sqrt{3}} \left( E_{i+1,1}^{n+1} - E_{i,1}^{n+1} \right) \right) \\
          + \varepsilon^2 \frac{\tilde{\Delta t}}{2 \Delta x} \alpha_{i-\frac{1}{2}} \left( F_{i, 1}^{n+1} + F_{i-1, 1}^{n+1} - \frac{c}{\sqrt{3}} \left( E_{i,1}^{n+1} - E_{i-1,1}^{n+1} \right) \right),
        \end{alignedat}
        \label{eq:formAlpha}
      \end{equation}
      whereas the asymptotic development of a standard discretization of \autoref{eq:diffusionEqua} would be
      \begin{equation}
        \begin{aligned}
          \varepsilon^2 \left(E_{i,0}^{n+1} + \rho c_v T_{i,0}^{n+1}\right) &= \varepsilon^2 \left(E_{i,0}^n + \rho c_v T_{i,0}^n \right)\\
          &+ \varepsilon \frac{c}{3} \frac{\Delta t}{\Delta x^2} \left( \frac{E_{i+1,0}^{n+1} - E_{i,0}^{n+1}}{\sigma_{i+\frac{1}{2}}} - \frac{E_{i,0}^{n+1} - E_{i-1,0}^{n+1}}{\sigma_{i-\frac{1}{2}}} \right).
        \end{aligned}
      \end{equation}
      So, we are looking for $\alpha_{i+\frac{1}{2}}$ such that the term of order $1$ in \autoref{eq:formAlpha} becomes a term of order $2$ with the expected coefficient of diffusion $\frac{c}{3 \sigma_{i+\frac{1}{2}}}$ and the term of order $2$ becomes a term of order $3$ and therefore negligible. In other words, we want the asymptotic development of $\alpha_{i+\frac{1}{2}}$ to be $\frac{2 \varepsilon}{\sqrt{3} \sigma_{i+\frac{1}{2}} \Delta x}$. One way to achieve this is to take 
      \begin{equation}
        \alpha_{i+\frac{1}{2}} = \cfrac{1}{1 + \sqrt{3} \sigma_{i+\frac{1}{2}} \frac{\Delta x}{2}}.
        \label{eq:asymptoticCorrectionP1}
      \end{equation}
      However, in the general case, we do not have $\lambda_{i+\frac{1}{2}}^+ = - \lambda_{i+\frac{1}{2}}^- = \frac{c}{\sqrt{3}}$. We can then replace \autoref{eq:asymptoticCorrectionP1} by
      \begin{equation}
        \alpha_{i+\frac{1}{2}} = \cfrac{1}{1- 3 \sigma_{i+\frac{1}{2}} \Delta x \cfrac{\lambda_{i+\frac{1}{2}}^+ \lambda_{i+\frac{1}{2}}^-}{c \left(\lambda_{i+\frac{1}{2}}^+ - \lambda_{i+\frac{1}{2}}^-\right)}}.
        \label{eq:asymptoticCorrectionDiffusiveLimit}
      \end{equation}
      Unfortunately, in numerical tests with $\sigma \Delta x$ close to $1$, the condition $f < 1$ is not preserved. Because $f$ is close to $1$ in this case, we write
      \begin{equation}
        \alpha_{i+\frac{1}{2}} = \cfrac{1}{1- 3 \sigma_{i+\frac{1}{2}} \Delta x \left( 1-f_{i+\frac{1}{2}}^2 \right) \cfrac{\lambda_{i+\frac{1}{2}}^+ \lambda_{i+\frac{1}{2}}^-}{c \left(\lambda_{i+\frac{1}{2}}^+ - \lambda_{i+\frac{1}{2}}^-\right)}}.
      \end{equation}
    We use $f_{i+\frac{1}{2}} = \frac{1}{2} \left( f_i^n + f_{i+1}^n \right)$ because numerical experiments have shown good results using this form. In the diffusion regime, because $F_{i,0}^{n+1} = 0$, we recover \autoref{eq:asymptoticCorrectionDiffusiveLimit}.

      Now that we have the form of $\alpha_{i+\frac{1}{2}}$, we can check that the proposed scheme is asymptotic preserving. We have
    \begin{equation}
      \alpha_{i+\frac{1}{2}} = \frac{2\varepsilon}{\sqrt{3} \tilde{\sigma}_{i+\frac{1}{2}} \Delta x} + \mathcal{O} (\varepsilon^2).
    \end{equation}
    Therefore,
    \begin{equation}
      \alpha_{i+\frac{1}{2}} \mathcal{F}_{i+\frac{1}{2}}^* = - \varepsilon \frac{c}{3 \tilde{\sigma}_{i+\frac{1}{2}}} \frac{E_{i+1,0}^{n+1} - E_{i,0}^{n+1}}{\Delta x} + \mathcal{O} (\varepsilon^2).
    \end{equation}
    We finally have
    \begin{equation}
      \begin{aligned}
        E_{i,0}^{n+1} + \rho c_v T_{i,0}^{n+1} &= E_{i,0}^{n} + \rho c_v T_{i,0}^{n} \\
        &+ \frac{c \tilde{\Delta t}}{3 \Delta x^2} \left( \frac{E_{i+1,0}^{n+1} - E_{i,0}^{n+1}}{\tilde{\sigma}_{i+\frac{1}{2}}} - \frac{E_{i,0}^{n+1} - E_{i-1,0}^{n+1}}{\tilde{\sigma}_{i-\frac{1}{2}}} \right).
      \end{aligned}
      \label{eq:discreteDiffusionEqua_appendix}
    \end{equation}
    \Autoref{eq:asymptoticDevDiscreteOrder0}, \autoref{eq:asymptoticDevDiscreteOrder1} and \autoref{eq:discreteDiffusionEqua_appendix} are standard discretization of \autoref{eq:asymptoticDevOrder0}, \autoref{eq:asymptoticDevOrder1} and \autoref{eq:diffusionEqua}, so this scheme is asymptotic preserving.
    \section{Expansion of \ion{H}{ii} region} \label{appendix:hii_region}

    In the test case described in \autoref{sect:hiiRegionExpansion}, some numerical noise appears, as a consequence of the long timescales. Let us recall that a time-implicit scheme is used, with large time steps for the radiative transfer. At each time step, the Newton-Raphson method is used and, at each iteration of this algorithm, an ill-conditioned linear system is solved, using an iterative process. This results in the appearance of some numerical noise.

    \begin{figure*}
      \begin{subfigure}[b]{0.5\textwidth}
        \begin{centering}
          \includegraphics[width=0.86\columnwidth]{hii_region}
          \caption[] {$4 \times 4$ MPI processes, ILU(k) factorization.}
          \label{fig:hii_region_prec_mpi:riluk_44} 
        \end{centering}
      \end{subfigure}
      \hfill
      \begin{subfigure}[b]{0.5\textwidth}
        \begin{centering}
          \includegraphics[width=0.86\columnwidth]{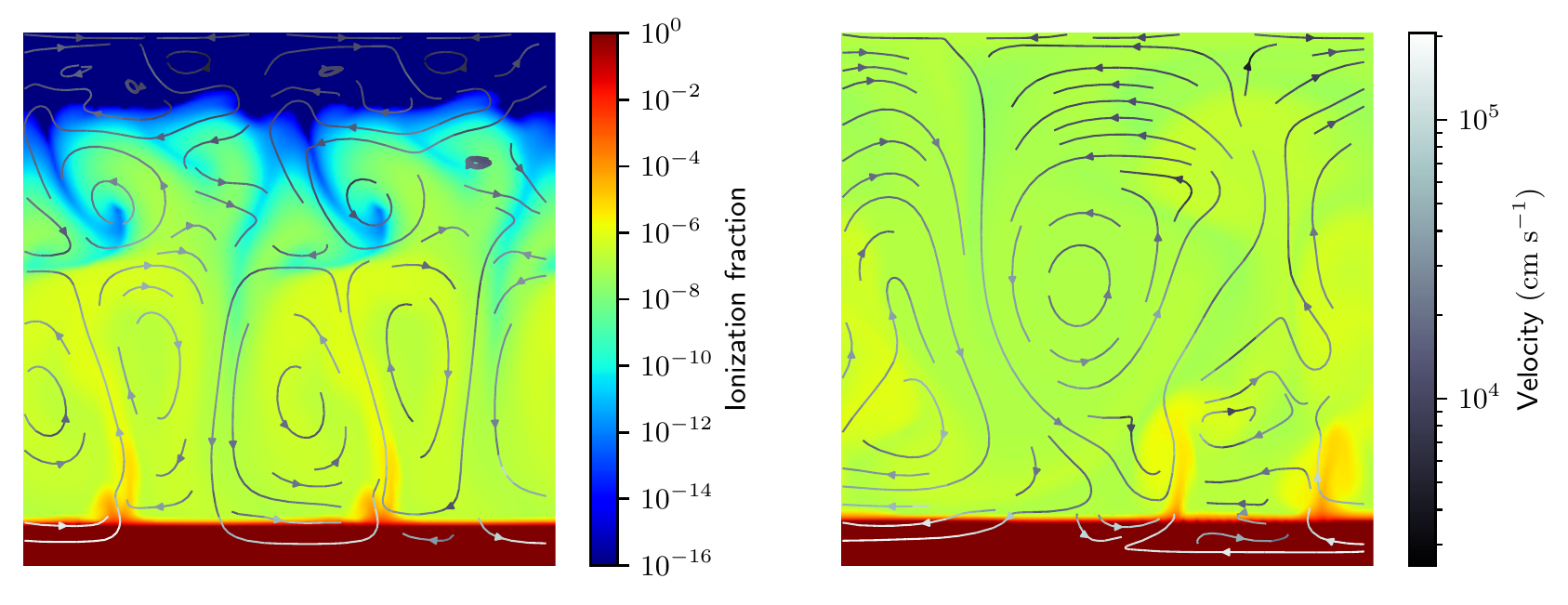}
          \caption[] {$2 \times 2$ MPI processes, ILU(k) factorization.}
        \label{fig:hii_region_prec_mpi:riluk_22} 
        \end{centering}
      \end{subfigure}
      \begin{subfigure}[b]{0.5\textwidth}
        \begin{centering}
          \includegraphics[width=0.86\columnwidth]{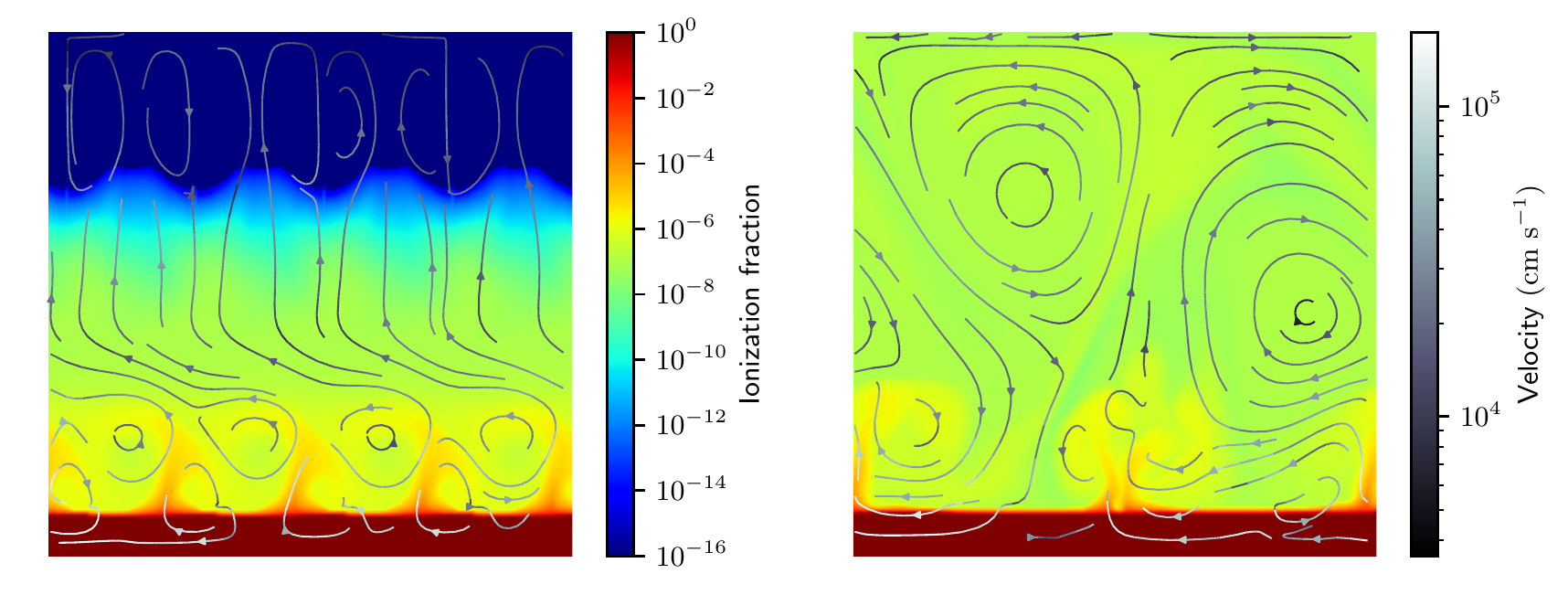}
          \caption[] {$4 \times 4$ MPI processes, Schwarz domain decomposition.}
          \label{fig:hii_region_prec_mpi:schwarz_44}
        \end{centering}
      \end{subfigure}
      \hfill
      \begin{subfigure}[b]{0.5\textwidth}
        \begin{centering}
          \includegraphics[width=0.86\columnwidth]{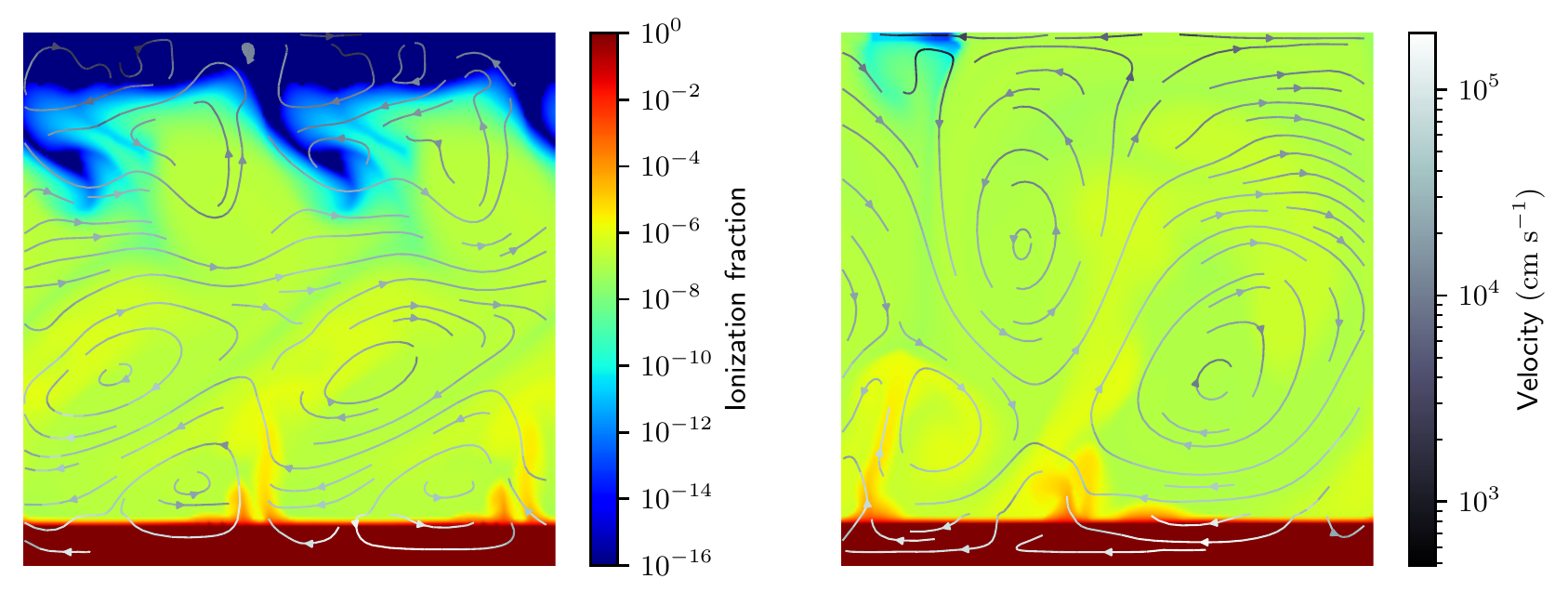}
          \caption[] {$2 \times 2$ MPI processes, Schwarz domain decomposition.}
          \label{fig:hii_region_prec_mpi:schwarz_22}
        \end{centering}
      \end{subfigure}
      \caption[]{Snapshots of the fraction of ionization and the velocity field at the final time $t_f = 10^{10}\ \mathrm{s}$ without the initial velocity perturbation (left panel) and with it (right panel). The physical domain is distributed across different numbers of MPI processes and different preconditioners have been used. \Autoref{fig:hii_region_prec_mpi:riluk_44} is the same figure as \autoref{fig:hii_region}.}
        \label{fig:hii_region_prec_mpi}
    \end{figure*}

  We have performed the same simulations as in \autoref{sect:hiiRegionExpansion} with different numbers of MPI processes and different preconditioners. The physical domain is either distributed over $4 \times 4$ MPI processes (\autoref{fig:hii_region_prec_mpi:riluk_44} and \autoref{fig:hii_region_prec_mpi:schwarz_44}) or $2 \times 2$ MPI processes (\autoref{fig:hii_region_prec_mpi:riluk_22} and \autoref{fig:hii_region_prec_mpi:schwarz_22}). We have also tried two preconditioners which allowed us to reach the final time with reasonable computational time: a standard ILU(k) factorization (\autoref{fig:hii_region_prec_mpi:riluk_44} and \autoref{fig:hii_region_prec_mpi:riluk_22}) and an additive Schwarz domain decomposition (\autoref{fig:hii_region_prec_mpi:schwarz_44} and \autoref{fig:hii_region_prec_mpi:schwarz_22}).

    \Autoref{fig:hii_region_prec_mpi} shows snapshots of the fraction of ionization and the velocity field at the final time $t_f = 10^{10}\ \mathrm{s}$. The shape of the small structures produced by the numerical noise varies with the number of MPI processes and the preconditioner. Furthermore, the propagation of the ionization front creates some velocity that also depends on the number of MPI processes and preconditioners. However, the position of the ionization front is not affected by these parameters.
  \end{appendix}

  \end{document}